\documentclass[%
 aip,
 amsmath,amssymb,
 reprint,%
]{revtex4-1}

\usepackage{graphicx}
\usepackage{subcaption}
\usepackage{dcolumn}
\usepackage{bm}

\usepackage[utf8]{inputenc}
\usepackage[T1]{fontenc}
\usepackage{mathptmx}
\usepackage{etoolbox}

\makeatletter
\def\@email#1#2{%
 \endgroup
 \patchcmd{\titleblock@produce}
  {\frontmatter@RRAPformat}
  {\frontmatter@RRAPformat{\produce@RRAP{*#1\href{mailto:#2}{#2}}}\frontmatter@RRAPformat}
  {}{}
}%
\makeatother
\begin{document}

\preprint{AIP/123-QED}

\title[]
{Spatio-temporal evolution of emission and absorption signatures in a laser-produced plasma}
\author{S.S. Harilal}
 \email{hari@pnnl.gov}
\affiliation{Pacific Northwest National Laboratory, Richland, Washington 99352, USA}%

\author{E.J. Kautz}
\affiliation{Pacific Northwest National Laboratory, Richland, Washington 99352, USA}%
 
\author{M.C. Phillips}%
\affiliation{James C. Wyant College of Optical Sciences, University of Arizona, Tucson, Arizona 85721, USA}

\date{\today}

\begin{abstract}

We report spatio-temporal evolution of emission and absorption signatures of Al species in a nanosecond (ns) laser-produced plasma (LPP). The plasmas were generated from an Inconel target,which contained  $\sim $ 0.4 wt. \% Al, using 1064 nm, $\approx$ 6 ns full width half maximum pulses from an Nd:YAG laser at an Ar cover gas pressure of $\approx$ 34 Torr. The temporal distributions of the Al I (394.4 nm) transition were collected from various spatial points within the plasma employing time-of-flight (TOF) emission and laser absorption spectroscopy and they provide kinetics of the excited state and ground state population of the selected transition. The emission and absorption signatures showed multiple peaks in their temporal profiles, although they appeared at different spatial locations and times after the plasma onset. The absorption temporal profiles showed an early time signature representing shock wave propagation into the ambient gas. We also used emission and absorption spectral features for measuring various physical properties of the plasma. The absorption spectral profiles are utilized for measuring linewidths, column density and kinetic temperature while emission spectra were used to measure excitation temperature.  A comparison between excitation and kinetic temperature were made at various spatial points in the plasma. Our results highlight that the TOF measurements provide a resourceful tool for showing the spatio-temporal LPP dynamics with higher spatial and temporal resolution than is possible with spectral measurements, but are difficult to interpret without additional information on excitation temperatures and linewidths.  The combination of absorption and emission TOF and spectral measurements thus provides a more complete picture of LPP spatio-temporal dynamics than is possible using any one technique alone.

\end{abstract}

\maketitle

\section{\label{sec:intro}Introduction}
Laser-produced plasmas (LPPs) are used in numerous applications ranging from analytical to inertial confinement fusion, and plasma diagnostics plays an important role in understanding the fundamental properties of the LPP as well optimizing their properties for various applications.  Compared to other man-made plasmas in the laboratory (e.g., steady-state discharge plasmas), there are several challenges in analyzing complex LPPs due to its transient nature and spatial inhomogenity. For example, the fundamental properties of an LPP change rapidly with space and time over several orders in magnitude.\cite{2018-APR-Hari} Hence, for comprehensive characterization of an LPP, the selected diagnostic tool should possess high time and space precision.  In addition to these, because of the existence of large gradients in temperature and density, multiple diagnostic tools are required to monitor the plasma physical conditions at various times  of its evolution (or distances from the target) due to sensitivity issues related to the selected diagnostic tool.\cite{Hutchinson2005} 

Among the plasma diagnostic methods, optical emission spectroscopy (OES) is the most utilized tool for LPP characterization and this can be related to its experimental simplicity as a non-intrusive technique.\cite{Aguilera2004, Hermann2017, gurlui2008experimental}   Besides, the combination of LPP and OES is the basis of well-known analytical tool laser-induced breakdown spectroscopy (LIBS) which is an established technique for several applications.\cite{Singh2020, Musazzi2014Book} Hence a large amount of works are available in the literature for the characterization of the LPP using OES.\cite{2020-AC-Liz, chen2015comparison, rao2016femtosecond} However, there exists certain limitations for using OES for LPP characterization, most prominent being that the LPP emits line radiation only within a certain temporal window during its evolution. At very early times of LPP evolution, the free-free and free-bound radiation dominate over bound-bound transitions and hence OES may not be a good tool for LPP characterization. Similarly, at later times of LPP evolution, the temperature of the plasma drops significantly so that the  electronic excitation process becomes weak or nonexistent. 

Active sensing methods such as laser absorption spectroscopy (LAS) and laser induced fluorescence (LIF) spectroscopy, which utilize the ground or lower level population, are useful for measuring properties of an LPP at later times of its evolution where the electronic excitation is not favored.\cite{Miyabe2015SCABLIF, Merten2018Pseudo, 2020-SCAB-Hari-Uhfs, duffey1995absorption, smith1998laser, Weerakkody2021, 2021-SCAB-Weeks-DCS, 2021-SCAB-Nicole, bushaw2009isotope, merten2018massing, 2021-SCAB-LIZ-Review-UO}  Although absorption spectroscopy is a well-established technique for gas sensing, its use for LPP characterization is limited and it is partly due to its active nature that necessitates the use of a light source such as laser,\cite{2021-PRE-Hari} arc lamp,\cite{Weerakkody2021} or frequency combs \cite{2021-SCAB-Weeks-DCS} for probing the absorption by the LPP species.  The properties of the probe light source also dictate the experimental methodology as well as information gathered. For example, AS employing tunable lasers provides extremely high spectral resolution, but its spectral band is limited to the scanning range of the laser system used and hence it is a highly a selective method.\cite{Miyabe2013Appl, Smith1999, 2021-PSST-Hari} Instead, the AS employing arc lamps provides broadband capability, but the spectral resolution available in this method is limited to the resolution of the detection system (e.g., spectrograph) which is similar to emission spectroscopy.\cite{Weerakkody2021} AS employing frequency combs, which is a relatively new technique, provides both spectral resolution and broadband capability.\cite{2021-SCAB-Weeks-DCS, 2021-OL-Kane-BurstDCS} 

The major aim of this work is to evaluate and compare the spatio-temporal evolution of both excited and ground state populations in an LPP by combining OES and LAS. Since emission and absorption methods are complementary, by combining these two techniques,  fundamental properties of the plasma can be measured at early and late times of LPP evolution as well as from various spatial positions. For example, the emission predominates at early times when the plasmas are hotter, whereas AS is better suited for LPP characterization when the plasmas cool down at later times. Similarly, thermal excitation predominates at closer positions to the target in an LPP, while cooler conditions exist at farther distances from the target,  but both the atomic number density and temperature distributions evolve in time after formation of the LPP. The spatial analysis of LPP was reported extensively in the literature using emission spectroscopy.\cite{Aguilera2004, 2020-AC-Liz, Wainright-SCAB-2021, FED-Khare-2021} However, nearly all previously reported absorption spectroscopic studies of LPPs to date were carried out at a certain distance from the target and the selection of the distance was based on optimizing the signal to noise ratio (SNR).\cite{weerakkody2020time, 2021-SCAB-Weeks-DCS, 2017-SR-Mark}    To the best of the authors' knowledge, no studies exist on the spatial mapping of the ground state population of an LPP system with high temporal precision and a direct comparison to the excited state population.  We measured optical time of flight (TOF) emission and absorption signatures at various spatial points in the plasma to investigate the population kinetics of excited and ground states of an Al atomic transition. The time-resolved emission and absorption spectral features are also monitored at various distances from the target surface to infer plasma properties.    

\section{\label{sec:methods}Experimental Methods}
\begin{figure}
\includegraphics[width=0.45\textwidth]{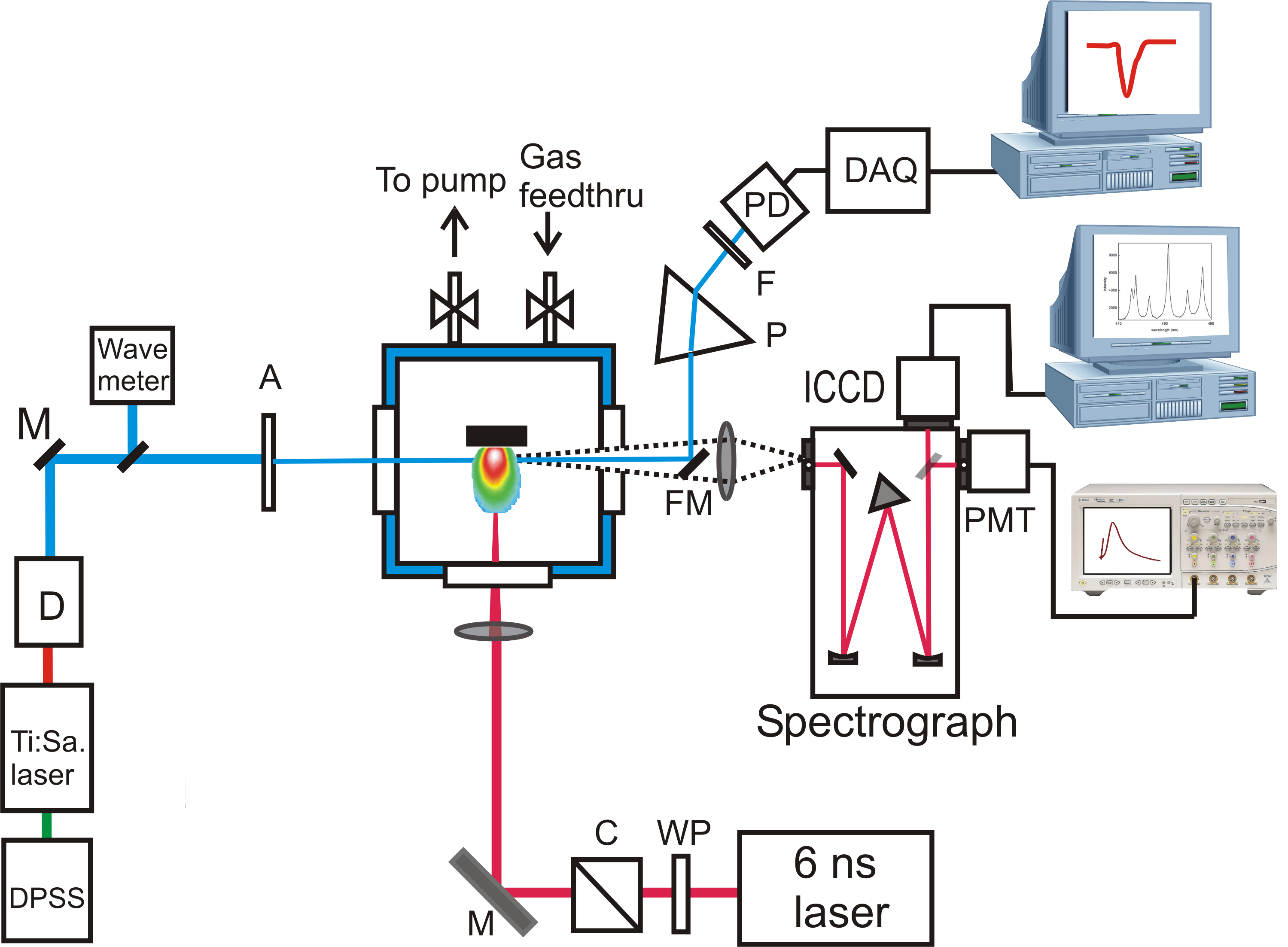}
\caption{ \label{expsetup} Experimental set-up for LPP-LAS and LPP-OES.  Acronyms given in the schematic are defined as follows: M: mirror, FM: folding mirror, A: aperture, DPSS: diode pump solid-state laser, D: frequency doubler, P: prism; F: filter, PD: photodiode, DAQ: data acquisition unit, ICCD: intensified CCD, C: cube polarizer, WP: waveplate, PMT: Photomultiplier tube} 
\end{figure}

A schematic of the experimental set-up is given in Fig. \ref{expsetup}. The plasmas were produced on an Inconel alloy target which contained 0.4 wt. \% Al using 1064 nm, 6 ns Full Width Half Maximum (FWHM) pulses from an  Nd:YAG laser (Continuum, Surelite III). The laser beam was attenuated using a combination of a half-wave plate and a cube polarizer. A f = 15 cm lens was used for focusing. The laser energy and spot size used were $\approx$ 50 mJ and $\approx$ 0.9 mm and the corresponding laser fluence and power density were $\approx$ 10 J/cm$^2$ and $\approx$ 2 GW/cm$^2$.  The repetition rate of the laser was 10 Hz. The Inconel target was positioned in a vacuum chamber, which contained glass/quartz windows for ablation and probe laser entrance. A motorized x-y-z translation stage was holding the the target chamber so that fresh target surfaces can be exposed  during ablation events. The experiments were performed in $\approx$ 34 Torr flowing ($\approx$ 3 l/min) Ar gas.      

For performing LAS, a frequency-doubled tunable Ti:Sapphire laser (M-Squared, Solstis and Solstis ECD-X) was used as the probe, which possessed a linewidth $ \leq $ 100 kHz. The probe laser with $\approx$ 0.75 mm diameter was directed through the plasma.  The frequency of the probe laser was monitored using a wavemeter (HighFinnesse WS-6).  Al I 394.401 nm (3s$^2$4s$^2$S$_{1/2}$ → 3s$^2$3p$^2$P$_0^{1/2}$, 0 - 25,347.756 cm$^{-1}$, log(gf) = -0.623)  transition is selected for this study. All spectroscopic information of the selected Al I transition is well documented in the literature.\cite{NIST} This line provided excellent emission and absorbance SNR at various times during the plasma evolution as well as from various spatial points in the plasma.   The transmission signal was measured using a a Si photodiode with 200 ns rise time (Thorlabs PDA36A). A combination of a prism and several bandpass and absorptive filters were used for filtering the spontaneous emission from the plasma and for attenuating the probe beam reaching the detector. A 16-bit analog-to-digital converter (ADC, National Instruments) at a 2 MHz sampling rate was used for signal digitization. The real-time analysis of the time-resolved absorbance and absorption spectrum were monitored using a LabVIEW program. For recording the absorption spectrum, the probe laser wavelength is stepped across the selected Al transition over $\approx$ 25 GHz. The time of flight absorbance was recorded using a 1 GHz oscilloscope when the probe laser wavelength is kept at the peak of the selected transition. The spatial analysis was performed by moving the chamber and the laser focusing lens in the z-direction by maintaining the probe beam alignment. 

For carrying out OES, the spontaneous light emitted from the plasma is collected and imaged onto the slit of a 0.5 m Czerny-Turner spectrograph (f/6.5, Acton Spectrapro 2500i) using an optical system consisting of two plano-convex lenses.  The spectrometer system consists of two detectors – an intensified CCD (ICCD, Princeton Instruments PiMAX4) - a multi-channel detector with 1024 $\times $ 1024 pixels (pixel size $\approx$ 13 $\mu$m $\times$ 13 $\mu$m) for recording the broadband spectrum, and a single channel photo-multiplier tube (PMT, Hamamatsu R955, rise time $\approx$ 2.2 ns) for measuring the temporal profiles of emission signal. A divertor mirror was used to change the  optical path of the wavelength-dispersed signal to ICCD or PMT. The spectral resolution available for the spectrometer was $\approx$ 0.04 nm with the use of a 2400 g/mm  holographic grating.  The spectrometer-PMT combination was used to record the kinetic distribution of selected Al atomic transition in the plume. For obtaining spatially resolved optical time of flight (OTOF) data, the plasma plume from different regions is imaged onto the spectrometer slit, and the PMT signal is fed to a 1 GHz oscilloscope. 

\section{\label{sec:results}Results}
Spatially and temporally resolved analyses of various species in an LPP are very important for understanding its kinetics. We analyzed the kinetics of excited and lower levels populations of a selected Al atomic transition at various distances from the target by combining the OES and LAS. The temporal distribution of a species in the plume at a certain distance from the target surface, commonly referred to as time-of-flight (TOF) emission\cite{thomas2020observation, Druffner2005, Skocic-TOF-2020} or absorption signal\cite{tarallo2016bah, AS1999-YANG} provides information about the persistence of the chosen species in the plume and the peak arrival/delay time at the selected spatial position and hence the expansion velocity. 

\begin{figure}[t]
\includegraphics[width=0.45\textwidth]{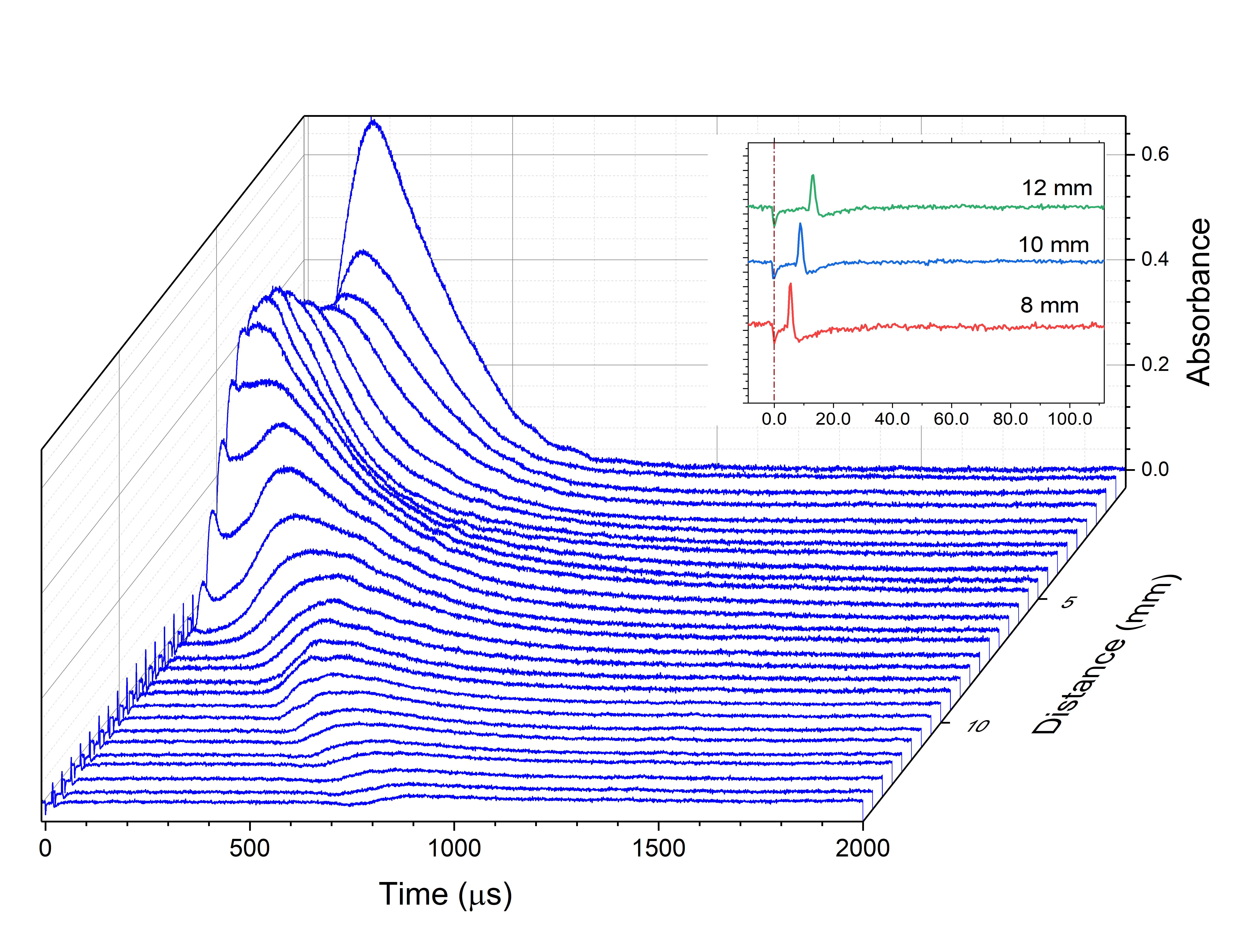}
\caption{\label{LAS-TOF} LAS-TOF at various distances from the target. Inset provides examples of absorbance signal early time evolution showing a dip at the plasma onset (dotted line) and a propagating wave.}
\end{figure}

For performing the spatial analysis of lower-level population, time-resolved probe transmission was recorded at various distances from the target surface when the probe laser was fixed at the peak of the Al I transition at $\lambda_0$ = 394.4 nm.  The recorded probe transmission $I(\lambda_0,t)$ was then converted to absorbance using the relation  $A(\lambda_0,t) = -\ln[I(\lambda_0,t)/I_0(\lambda_0)]$ where $ I_0 (\lambda_0)$ is the transmitted intensity before the LPP onset.   The measurements were averaged over six laser shots, and the results are given in Fig. \ref{LAS-TOF}.

\begin{figure}[t]
\includegraphics[width=0.45\textwidth]{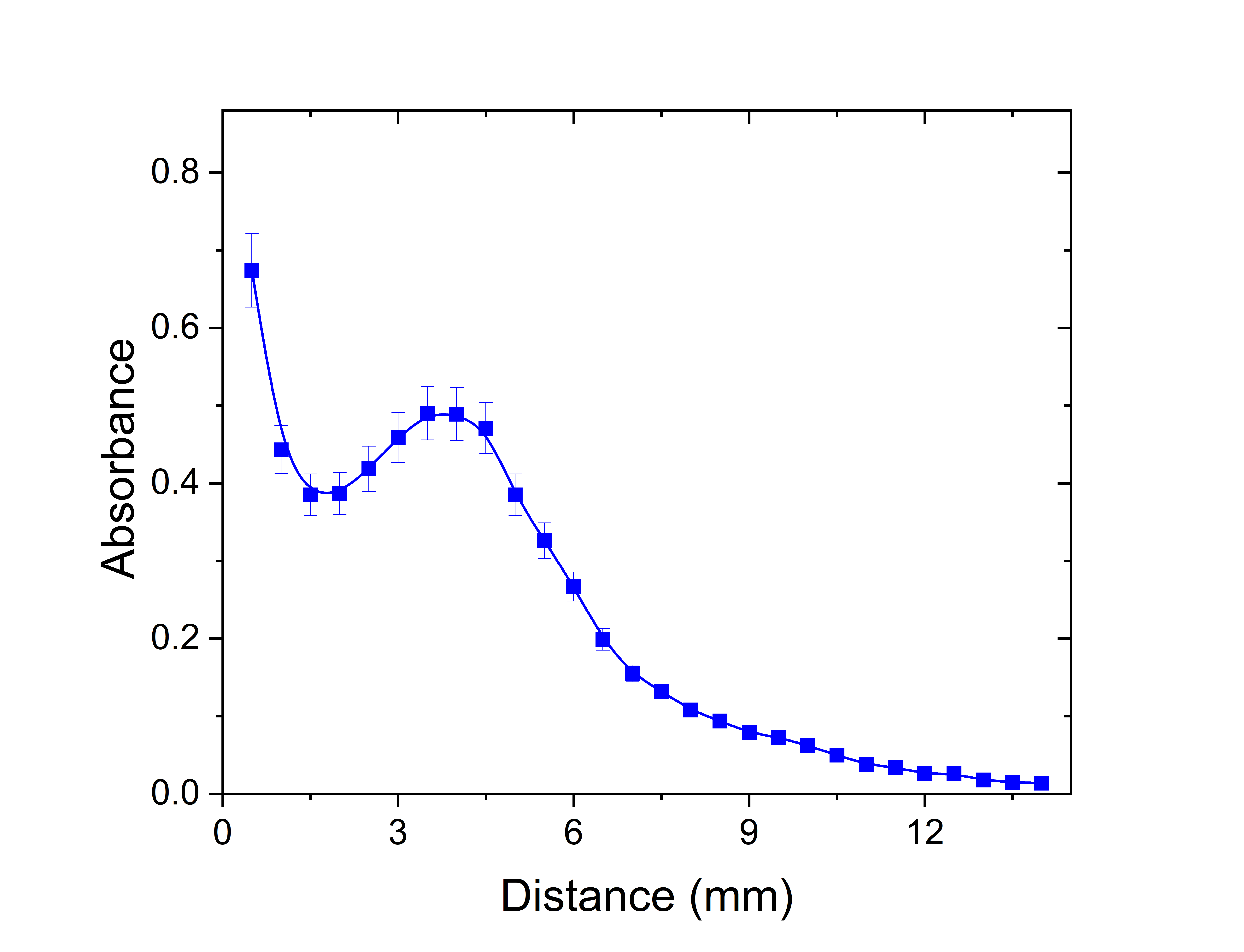}
\caption{\label{LAS-ATP} The peak absorbance signal at various distances are given.}
\end{figure}

The absorption temporal profiles showed significant variation with respect to the distance from the target. The maximum peak absorbance is observed closer to the target. The absorbance temporal profiles showed a sharp 'dip' immediately after the laser-plasma onset which was present for all distances.  A short duration peak in the absorption is seen to appear soon after ablation, with arrival time dependent on distance from the sample, as shown in the inset to Fig. \ref{LAS-TOF}.  These features will be analyzed in the discussion section.  The maximum absorbance at various spatial positions in the LPP plumes is given in Fig. \ref{LAS-ATP}. The peak absorbance signal showed two maxima: closer to the target ($\approx$ 1 mm), and at $\approx$ 4 mm. The key points from the spatially and temporally resolved absorption profiles are as follows: (1) the maximum absorbance is observed closer to the target although its arrival time is delayed, (2) with increasing distance from the target, the absorbance signal is found to decrease up to $\approx$ 2 mm from the target, followed by a slow increase up to $\approx$ 4 mm and then decays with increasing distance, (3)  multiple peak structures in temporal profiles are evident at farther distances ($\geq$ 5 mm), (4) the LAS signal persistence is increased with distance and at farther distances LAS signals lasted more than 2 ms, (5) the absorbance signal is negligible at early times for distances $\geq $  7 mm. 

\begin{figure}
\includegraphics[width=0.45\textwidth]{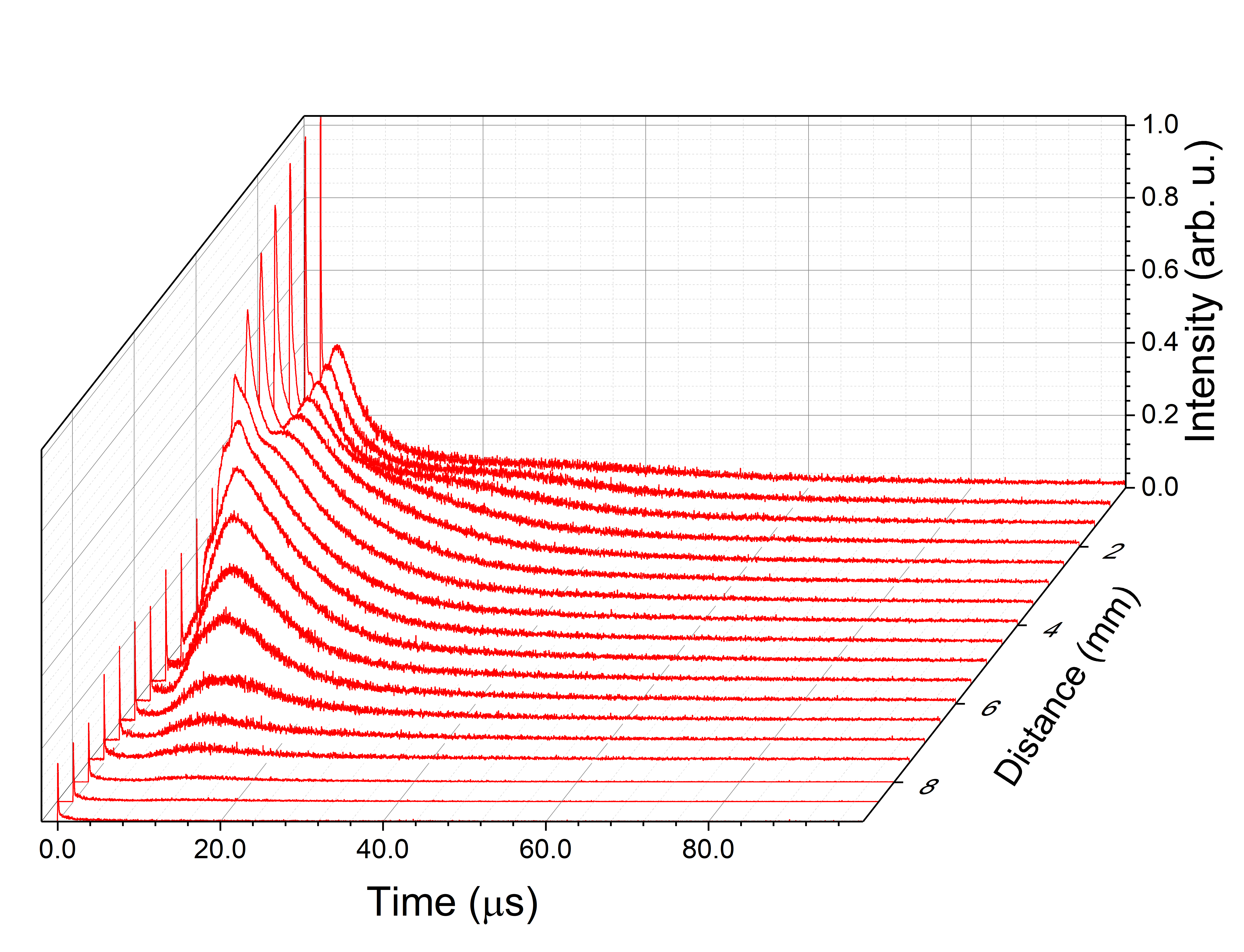}
\caption{ \label{OTOF} Emission TOF at various distances from the target. } 
\end{figure}

For monitoring the temporal profiles of emission signals at various spatial locations in the plasma, optical TOF emission profiles of Al I (394.4 nm) were recorded using the combination of the monochromator and PMT, with results given in Fig. \ref{OTOF}.  Such OTOF profiles provide the delay as well as the persistence of an emitting species in the plume and provide information on the excited level population of the selected transition.   All OTOF profiles show a prompt peak (sharp, at early times), followed by a sharp rise in signal intensity that is steepest at closer distances to the target. The OTOF profiles also showed multiple peak structures at shorter distances ($\leq$ 4 mm).  Some of the notable differences between the absorbance and emission signals are: (1) although absorbance signals are observed for distances up to $\approx$ 14 mm, the emission signal is found to be very weak after $\geq$ 8 mm; (2) the emission signal persisted much shorter time compared to absorbance signals, typically by an order in magnitude; (3) the multiple peaks in the TOF profiles are more apparent in emission signals compared to absorption signals.  

\begin{figure} 
\includegraphics[width=0.5\textwidth]{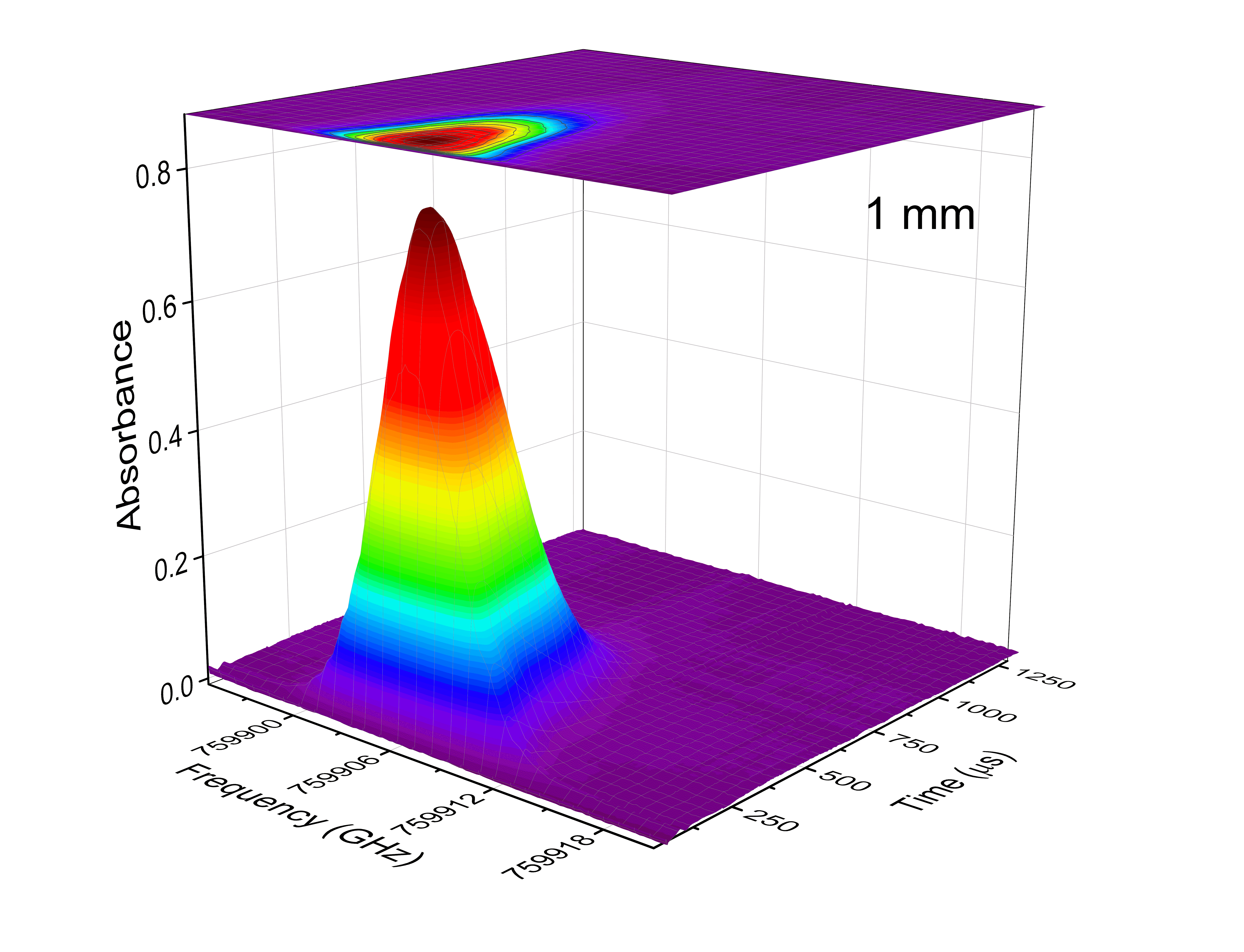}
\caption{\label{TRASfigure1} 2D and 3D contour showing TRAS for Al I transition at 394.401 nm recorded at a distance 1 mm from the target. The measurements were performed by stepping the tunable laser wavelength across the selected transition. Each spectrum corresponds to six laser shot averaging.}
\end{figure}

The emission and absorption TOF profiles give information on the temporal distribution of  the excited and ground state population  of the selected Al transition at various distances from the target. However, the TOF profiles are useful only for monitoring the relative populations, and it is not useful for inferring the plasma's fundamental properties (e.g., temperature, density). Hence, time and space resolved spectroscopic studies are carried out. An example of 2D and 3D time-resolved absorbance spectra (TRAS) contour plot is given in Fig. \ref{TRASfigure1}. The measurement was taken at a distance 1 mm from the target and by stepping the wavelength of the probe laser across the selected Al transition and 6 laser shots were averaged for each step. The peak absorbance at the line center was $ \approx$ 0.7.  

\begin{figure*}[hbt!]
\includegraphics[width=0.7\linewidth]{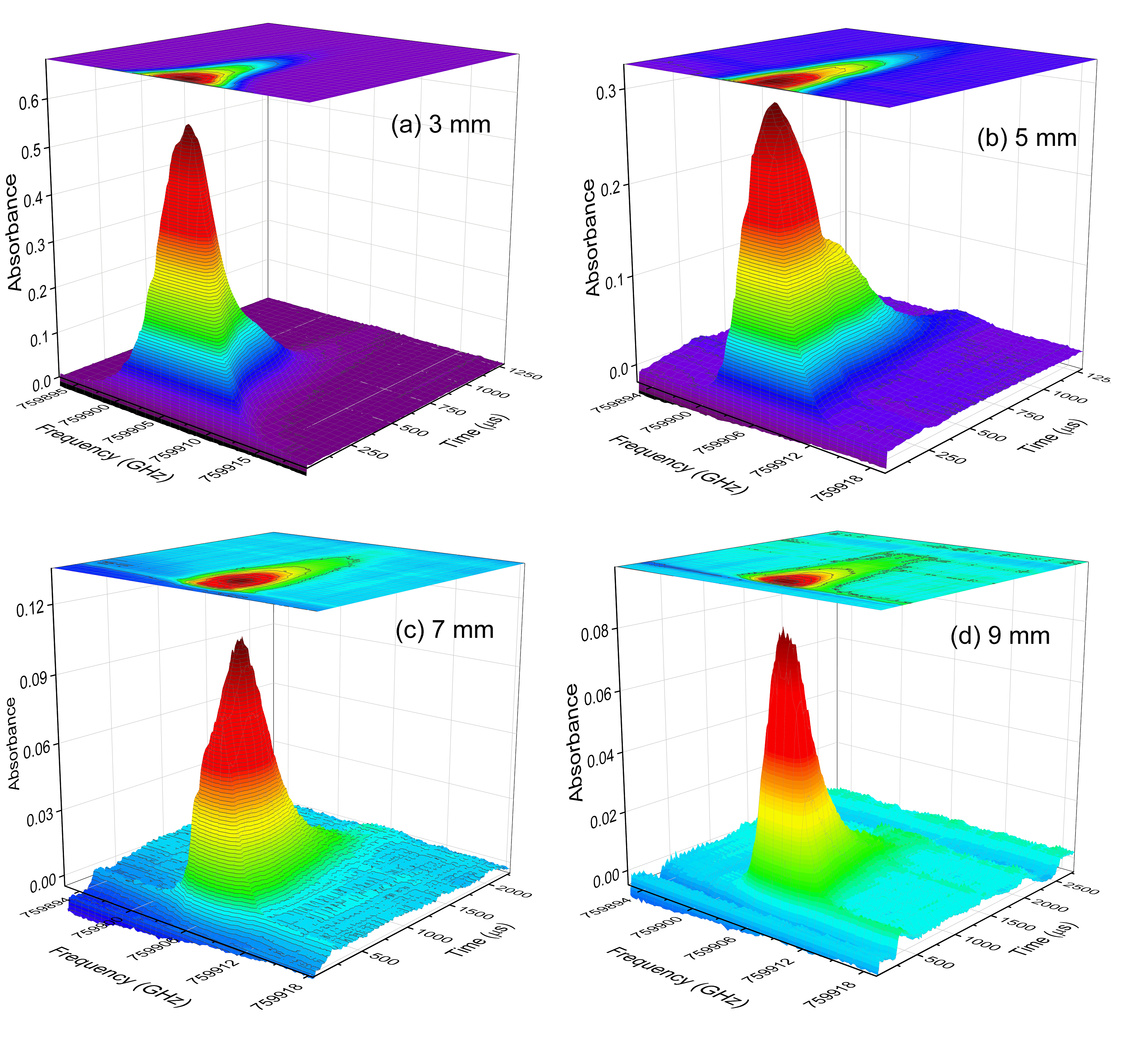}
\caption{\label{TRASfigure2} 2D and 3D contours showing TRAS for Al I transition at 394.401 nm for the following distances: (a) 3 mm, (b) 5 mm,
(c) 7 mm, and (d) 9 mm.}
\end{figure*}

The time-resolved absorption spectroscopy contours  recorded  at additional distances from the target are given in Fig. \ref{TRASfigure2}. Qualitatively the TRAS results show similar features as the  LAS-TOF results; namely, a delayed onset of absorption and a longer persistence of absorption at larger distances from the target.  It is also apparent that the spectral broadening changes with time, especially in regions closer to the target.  Additional information may be obtained by spectral fitting of the measurements and investigating how the fit parameters change with time and space.   

\begin{figure}
\includegraphics[width=\linewidth]{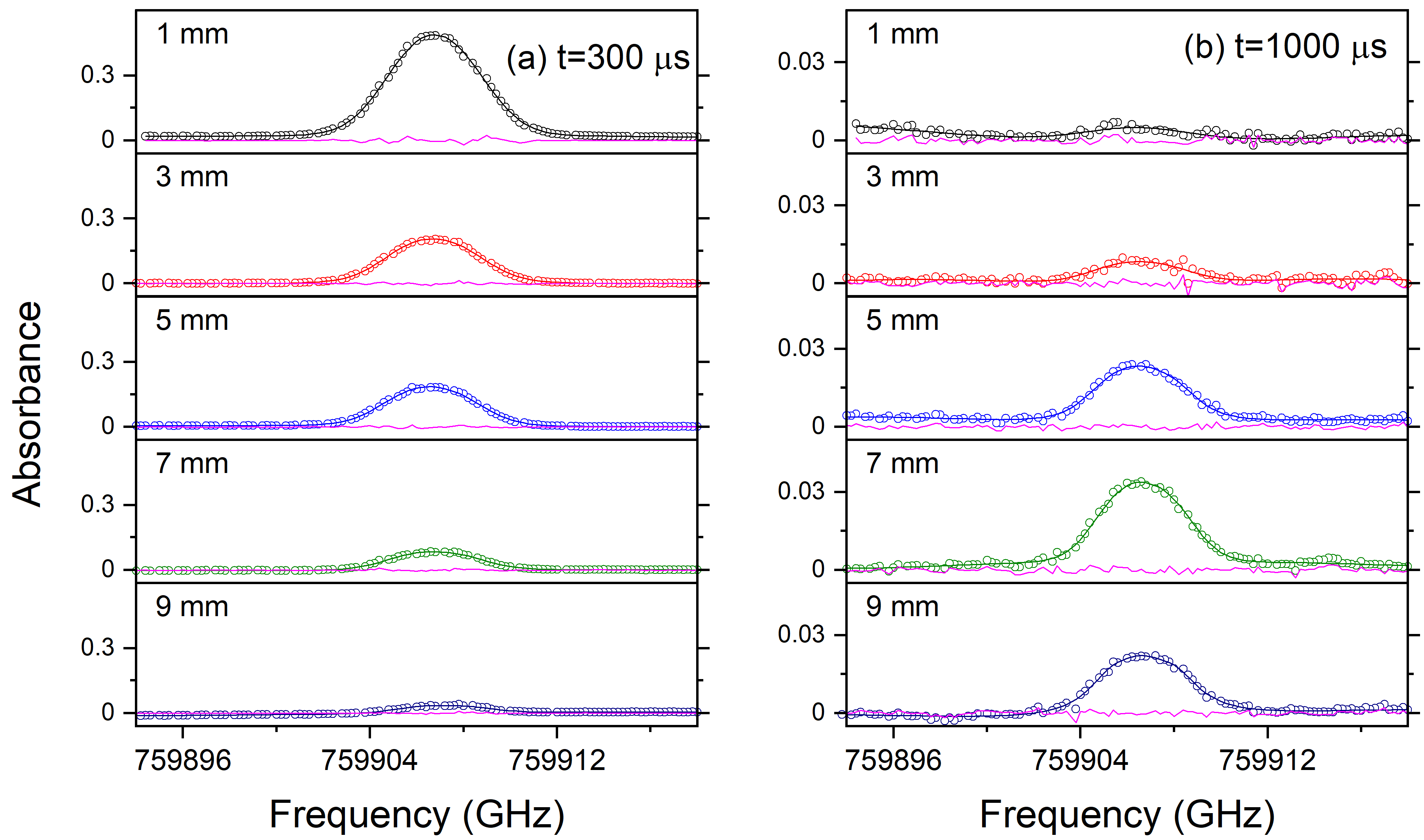}
\caption{\label{LAS-spectra} Examples of measured absorption spectra (open circles), best fit spectra (solid lines), and fit residuals (magenta lines) are given for various distances. (a) measured at 300 $\mu$s and (b) at 1000 $\mu$s.}
\end{figure}

The experimental spectra given in Figs. \ref{TRASfigure1} and \ref{TRASfigure2}  were fit using Voigt profiles for determining FWHM, peak areas, and Gaussian linewidth.  Examples of spectral fits are shown in Fig. \ref{LAS-spectra} for various distances from the target and for two select times (300 $\mu$s and 1000 $\mu$s).  The spectral fits also accounted for the hyperfine structure of the $^{27}$Al I transition, which was modeled using sums of Voigt functions with equal widths. Further details about the spectral fitting approach can be found elsewhere.\cite{2021-SCAB-Nicole}  The quality of the spectral fits is highly dependent on the SNR of the experimental data; therefore, the poor fits obtained at the extremes of distance/times were discarded due to low signals and high uncertainties. 

\begin{figure}[hbt!]
\includegraphics[width=0.8\linewidth]{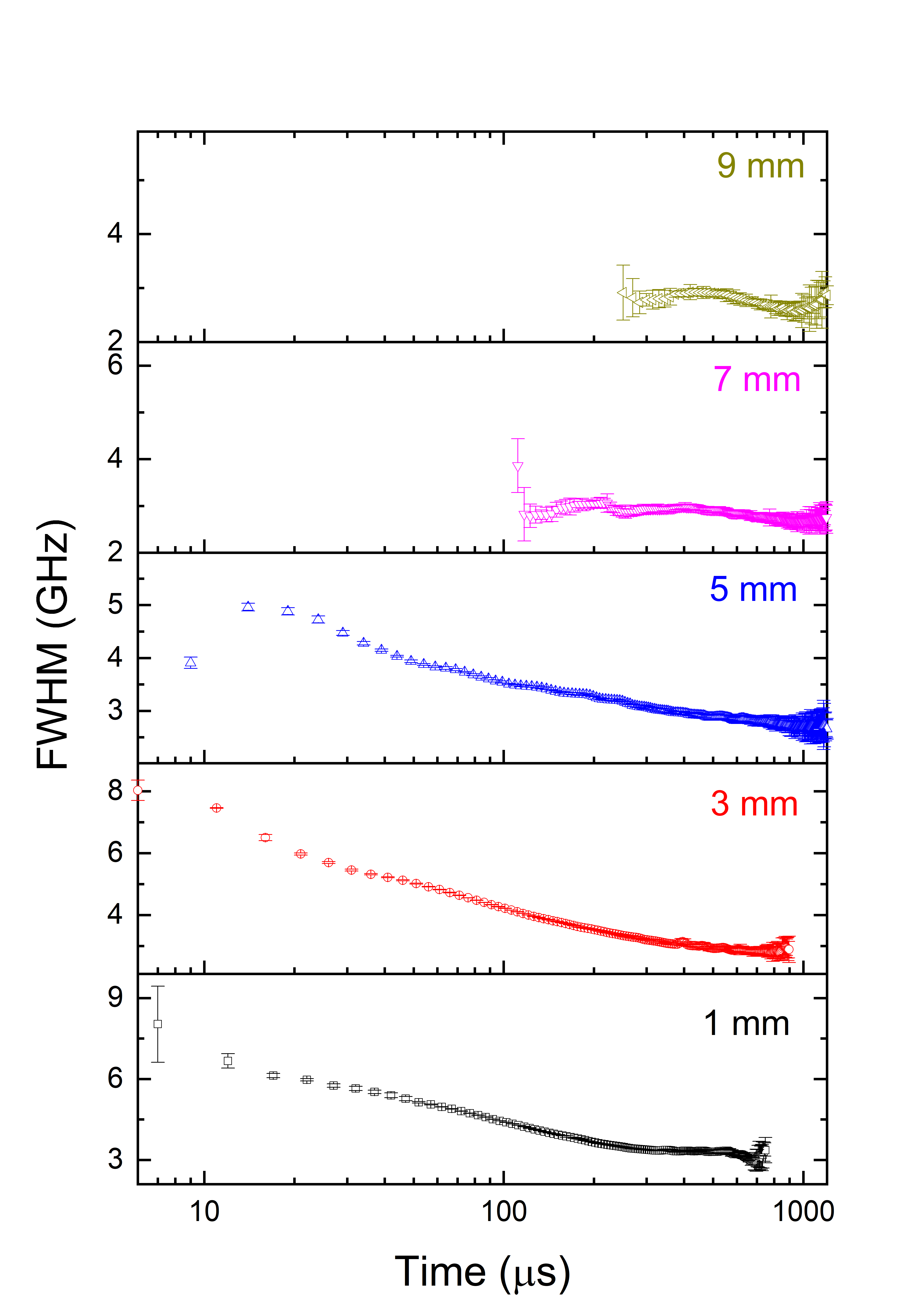}
\caption{\label{FWHM} The time evolution of linewidth (FWHM) for various distances from the target.}
\end{figure}

The time evolution of FWHM of Al I transition measured from the absorption spectra for various distances are given in Fig. \ref{FWHM} and the linewidth given is for each hyperfine transition. It must be noted that the experimentally measured linewidth is found to be broader than the FWHM given in Fig. \ref{FWHM} due to the overlap of the unresolved hyperfine transitions.   For shorter distances from the target ($\leq$ 5mm), the linewidth showed a reduction with time ($\leq$ 100 $\mu$s), and it leveled off at later times.  But, at farther distances (e.g., 7 mm and 9 mm), the FWHM showed insignificant variation.  The Voigt linewidths are approaching $\approx$ 3 GHz for all distances studied at times $\geq$ 600 $\mu$s.  The uncertainty in the spectral analysis is found to be high at the earliest and later times due to low absorbance levels.  

Under the assumption of spatially uniform conditions along the probe beam path, each absorbance spectrum given in Fig. \ref{LAS-spectra} is expressed as the product of an absorption coefficient $\alpha(\nu,t)$ and an optical path length $L : A(\nu,t) =\alpha(\nu,t)\times L$. Here the absorption coefficient is the product of the absorption cross-section with the difference in population density between lower and upper states of the probed transition, and is given by: $     \alpha_{ij}(\nu) =\tilde{\sigma_0} g_{i}f_{ij}  \left[\dfrac{n_i}{g_i}-\dfrac{n_j}{g_j}\right]  \chi(\nu) $ where $\tilde{\sigma_0} = e^2/4\epsilon_0 m_e c$ is a constant equal to $2.654 \times 10^{-6} $m$^2$s$^{-1}$, $i(j)$ denotes the lower (upper) level of the transition, $g_i$ is the level degeneracy of the $i$-th level, $f_{ij}$ is the transition oscillator strength, $n_i$ is the number density in the $i$-th level (m$^{-3})$, and $\chi(\nu)$ is a normalized lineshape function with unit area. For the LPP conditions measured, which follow  thermal distributions through Boltzmann statistics, $n_j << n_i$, in which case the above equation is simplified to: $A(\nu,t) = \alpha_{ij}(\nu,t)\times L = \tilde{\sigma_0} f_{ij} \chi(\nu,t) n_i(t)\times L$. According to the above equation, the spectrally integrated area of the absorbance peak is proportional to the path-integrated lower state number density or column density of the selected transition. Hence the  column density $n_i \times L$ for Al is calculated for each time step using the relation 
$\int{\alpha(\nu)  L  d\nu} = 2.654 \times 10^{-6}  f_{12} n_i L $, where the left-side of the equation is the spectrally integrated absorbance as determined from the peak areas.  The Gaussian FWHM contribution ($w_g$ (nm)) in the spectral profiles is used for measuring the kinetic temperature (T$_k$) using the relation  $   w_g = 7.16 \times 10^{-7}\lambda_0 (T_k/m)^{1/2} $
where  $\lambda_0$ is the wavelength in nm, $m$ is the mass of the species in amu and T$_k$ is in K.  Results on spatial and temporal dependence of column density and kinetic temperature are shown in section IV.

\begin{figure}[hbt!]
\includegraphics[width=0.9\linewidth]{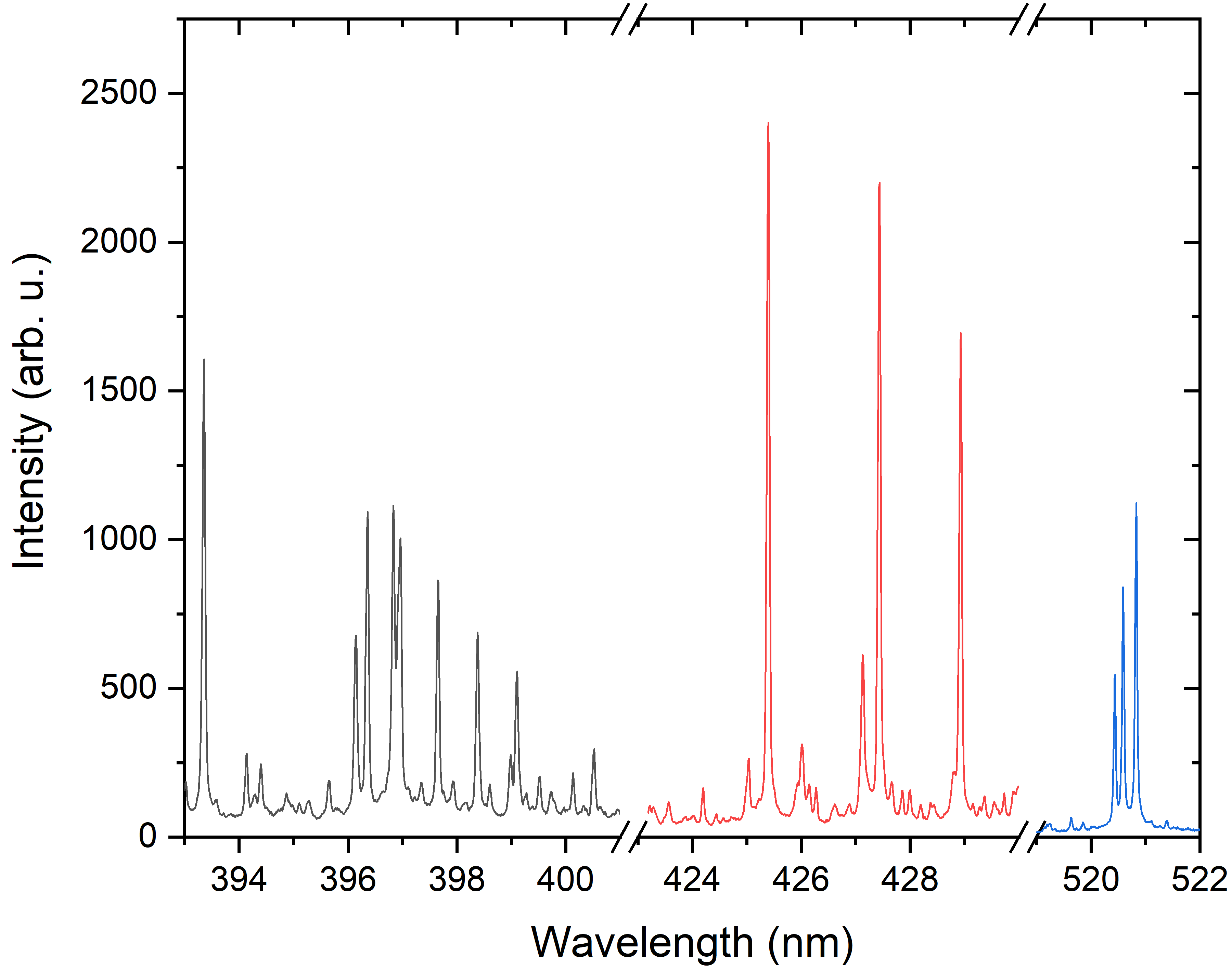}
\caption{\label{OES-spectrum} An example of Cr I spectral lines used for excitation temperature measurement is given. The given spectrum is recorded from a distance 1mm and using a gate delay/width of 1 $\mu$s/1 $\mu$s.}
\end{figure}

Emission spectral features were collected from various times after the plasma onset and distances from the target surface for measuring the excitation temperature ($T_e$). The Inconel target contains $\approx$ 19 \%  (by wt.) Cr and several Cr I transitions were used for this study. The Boltzmann plot method was used for measuring excitation temperatures using the following Cr I transitions: 394.15 nm, 396.37 nm, 398.39 nm,  399.11 nm, 400.13 nm, 520.451nm, 520.6 nm, and 520.84 nm. The spectroscopic constants for the lines were obtained from literature.\cite{NIST}  An example of optical emission spectral features collected at 1 mm from the target with a gate delay/width of 1$\mu$s/1$\mu$s is given in Fig. \ref{OES-spectrum}.  The time evolution of the excitation temperature measured at various distances from the target is presented in section IV.  

\section {Discussion} 
 
Optical TOF is a powerful technique for investigating velocity distribution of various species in the plasma and several authors previously used it for optimizing pulsed laser deposition (PLD) film quality and understanding plasma chemistry.\cite{Druffner2005, Ying2015, tarallo2016bah, 2021-SCAB-LIZ-Review-UO, thomas2020observation, Sakeek-1991-APL} Emission and absorption TOF profiles will also provide important parameters related to the evolution of laser ablated materials such as the arrival/delay time, velocity distribution, the signal intensity and persistence at various spatial positions in the plasma. In this article, a comprehensive evaluation of kinetics of excited and ground population of Al species at various spatial points in an LPP is made by combining emission and absorption TOF combined with analysis of TRAS and OES measurements. The transition selected for this study is a resonance line (Al I at 394.4 nm) and hence coupled to the ground level, allowing absorption to be detected for conditions of low excitation temperature at later times of plasma evolution. However, the Al transition selected for this study is prone to self-absorption.\cite{el2005evaluation} So, the contribution of self-absorption in the emission temporal profiles cannot be ruled out in the present experiment. Self-absorption will be absent in the absorption temporal profiles. 

The TOF profiles showed multi-component structures for both absorption and emission signatures and similar structures in the TOF profiles were reported previously in  absorption\cite{miyabe2012doppler, Cheung-JAP1991, Sakeek-1991-APL} and emission profiles\cite{thomas2020observation, 2001-ASS-Hari}.  The plume splitting phenomena in LPPs were reported extensively in the literature at moderate ambient pressure levels using 2D photography where two or more emission intensity maxima were observed and moved with different velocities. Some of the reported explanations given for plume splitting include  snow-plow effect and subsequent deceleration of plume front, multiple scattering of plume particles by the ambient medium,  ambipolar effect, and slowly propagating nanoparticles.\cite{JPD2021-Volkov, Bulgakov2000, focsa2017plume, PRL97-Geohegan}  

\begin{figure}[hbt!]
\includegraphics[width=0.8\linewidth]{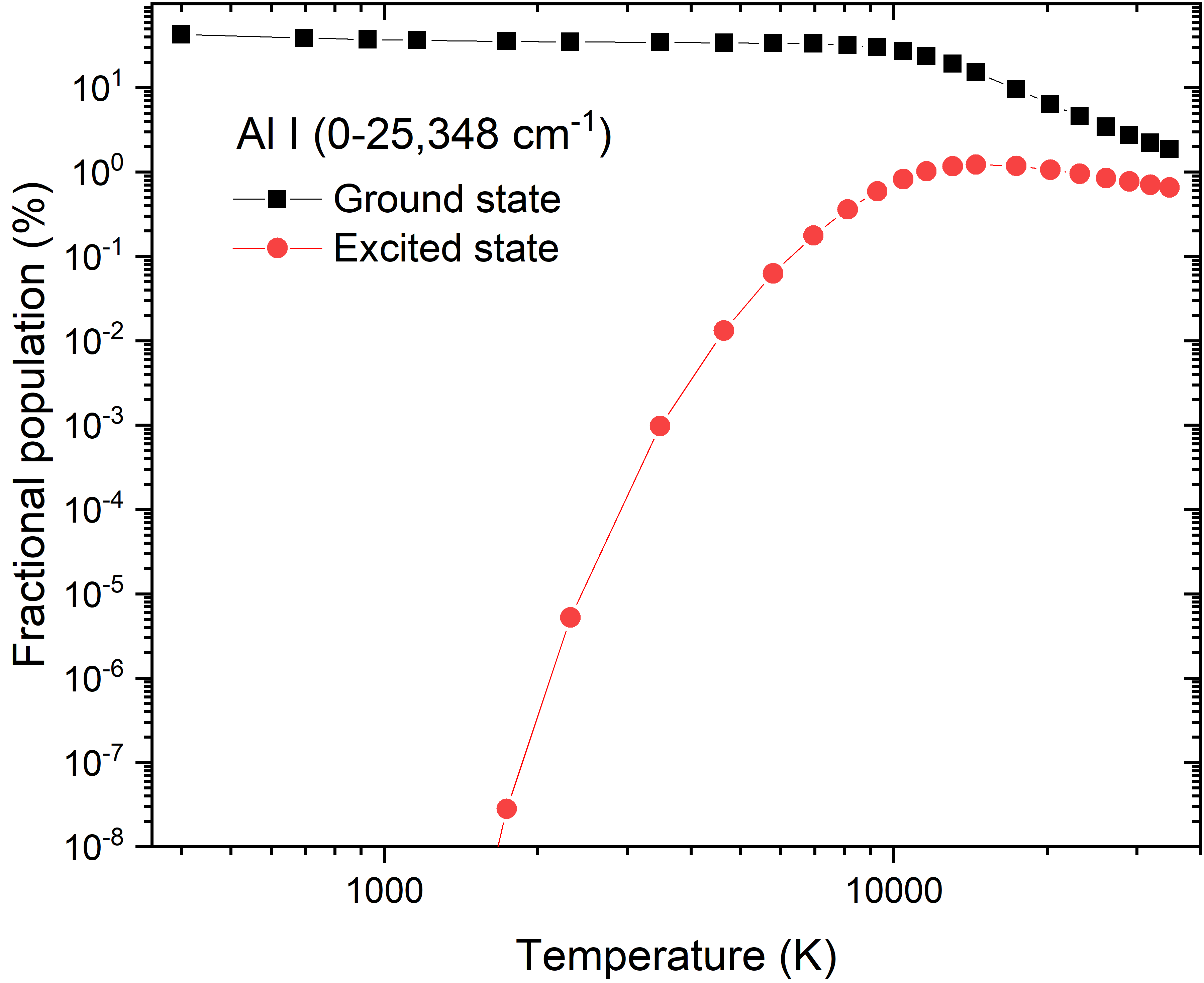}
\caption{\label{Boltzmann} The fractional population of upper and lower energy levels of Al I 394 nm transition based on Boltzmann equation.}
\end{figure}

 Many of the differences in temporal profiles and persistence of emission and absorption signals can be understood by considering excitation temperature, which determines the fractional populations of upper and lower states with temperature. Under local thermodynamic equilibrium (LTE) the level populations follow a Boltzmann distribution: $n_i=n_{tot}g_i Z(T_{ex})^{-1} exp(-E_i⁄kT_{ex})$, where  E, g and n$_{tot}$ are the energy, degeneracy and the total atomic number density, and $Z(T_{ex})= \Sigma g_i exp(-E_i / kT_{ex})$ is the partition function. The calculated fractional population versus excitation temperature for upper and lower energy levels of Al I 394.4 nm are given in Fig. \ref{Boltzmann}.  According to Fig. \ref{Boltzmann}, the ground state of Al transition will be populated at lower temperatures while higher plasma temperatures are required to populate the upper energy level of Al. A significant reduction in the lower state population is also noticeable when the temperature of the plasma system is $\geq$ 10,000K. Hence, under conditions of high excitation temperature the absorption signal from the ground state is reduced.  Likewise, for low excitation temperatures the emission signal from the excited state decreases rapidly.
 
Compared to absorption, the emission signals peaked early and decayed rapidly after the plasma onset.  The emission signal depends on the excited state population, and it requires electronic excitation, which happens only when the temperature of the plasma system is high enough (see Fig.\ref{Boltzmann}). According to Boltzmann distribution, the fractional population of the upper energy level becomes largest when the temperature is $\geq$ 10000 K.  The temporal absorption profiles were recorded at farther distances from the target (up to 14 mm), while the emission signals were found to be weaker for distances $\geq$ 8 mm.  These combined observations indicate that the  temperature of the plasma at farther distances and later times is too low to support significant excited state population, and thus most of the Al atoms are in the ground level.  As shown below, measurements of the excitation and kinetic temperatures support this conclusion.

The emission and absorption TOF profiles were line-of-sight averaged. It must be mentioned here that the LPP is an inhomogeneous medium, and significant variation in temperature and density exist in both expanding directions viz.  orthogonal and parallel to the target surface. Hence, the TOF profiles recorded at various spatial points represent the average excited and ground state distribution of Al atoms along the line-of-sight.   Both absorption and emission TOF profiles showed several structures in their temporal profiles, but it changed significantly with the time of observation. So a direct comparison between emission and absorption temporal profiles at various times during the plasma evolution is made for understanding its kinetics and relation to plume expansion. 

The following subsections discuss in more detail the spatio-temporal dynamics of emission and absorption TOF signals, as well as the column density and kinetic temperature determined from analysis of the TRAS measurements.  The discussion is divided into different temporal regions depending on the relevant dynamics of the LPP.  To better visualize the spatial evolution of the plume, the TOF axes given in Figs. \ref{OTOF} and \ref{LAS-spectra} are transposed.  For  comparison with the TOF results, the column density and kinetic temperatures determined from the spectral fits are plotted similarly to show spatial variations as a function of time.  Fit results are not plotted for spatial/temporal regions with poor fit quality due to low absorption levels.  In all plots, the lines show spline fits to the discrete data points to highlight trends in the spatial dependence of the plume with time.

It is also noted that the emission TOF profiles have smaller time steps compared to the absorption TOF profiles due to the differences in the data collection sampling rate (50 MHz vs 2 MHz). The response time of the detector used for emission (PMT) was faster ($\approx$ 2 ns) compared to the  detector used for absorption measurement ($\approx $ 200 ns). Likewise, the time steps for the TRAS results are larger (500 ns) due to differences in data acquisition hardware for the spectrally-resolved measurements.

\subsection{Temporal evolution 0-2 $\mu$s}
\begin{figure}[hbt!]
\includegraphics[width=0.8\linewidth]{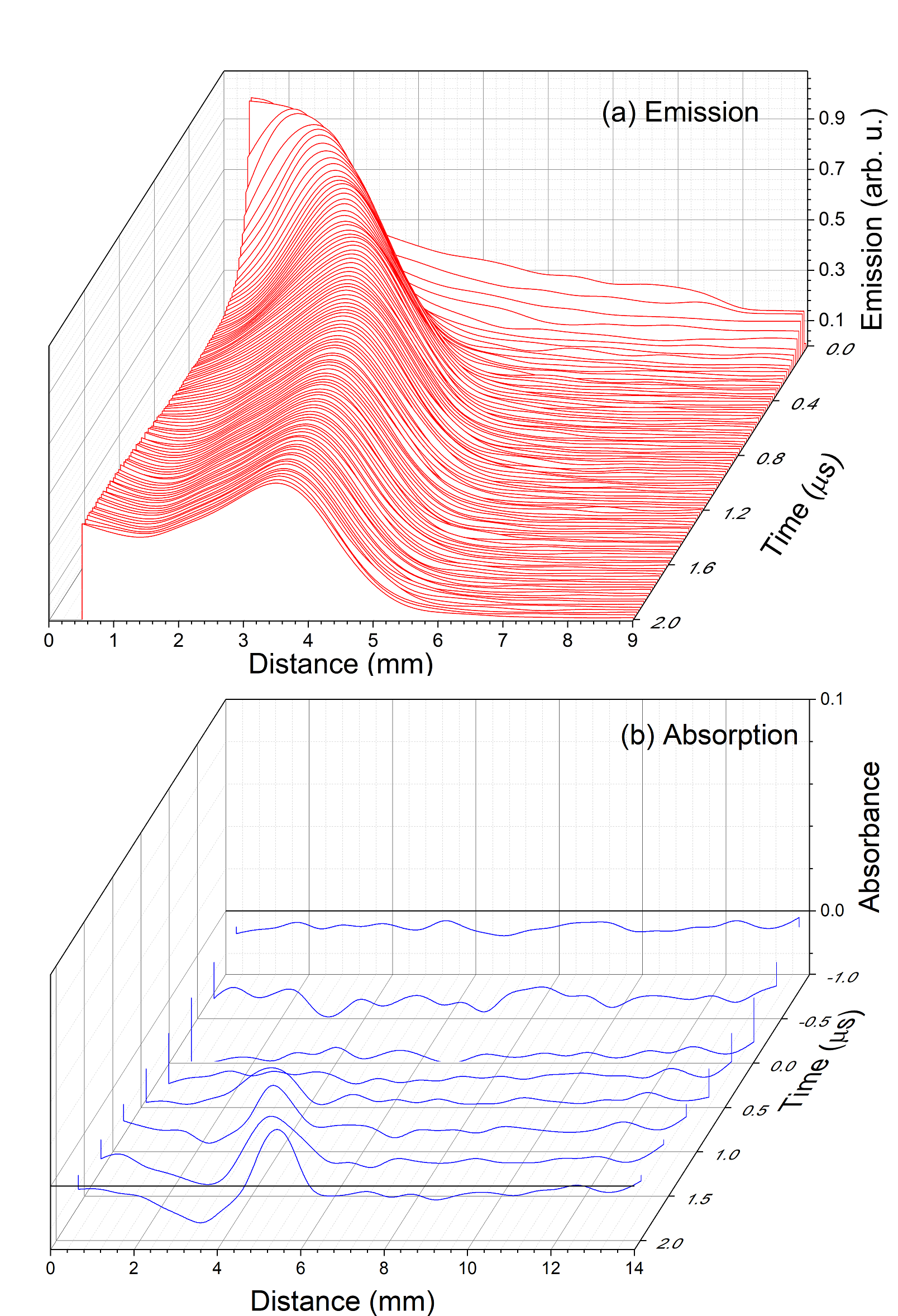}
\caption{\label{EA-TOF-0-2us} The spatio-temporal evolution of emission and absorption signals for times $\leq$ 2 $\mu$s after the laser-plasma onset.}
\end{figure}

 The spatio-temporal evolution of emission and absorption signal profiles at times $\leq 2 \mu$s after the laser-plasma onset are given in Fig. \ref{EA-TOF-0-2us}. Note that the spatial scales given for emission and absorption signals are different.  The emission signal showed a prompt signature appearing immediately after ablation for all distances from the target and persisting for $\approx$ 200 ns. After the prompt signal decays, the emission signal shows a rapid expansion and formation of a spatially distinct plume structure, peaked at a distance of $\approx$ 3.5 mm from the target. 

The absorption TOF shows the immediate appearance of a negative absorption signal at all distances from the target, which is consistent with detection of a short-lived prompt emission signal during these times.  As the emission signal decays, a spatially-localized absorption signal appears and begins propagating away from the sample.  This localized absorption signal is consistent with formation of a shock-wave propagating away from the target, as detailed further in Sections IV B and C.  A negative absorption region (dip) is also apparent  behind the shock front. 
The measured absorption spectra at these times (not shown) showed no evidence of resonant absorption due to Al atoms.  However, the the spectra did show a prompt negative-going absorption signal for all distances, appearing immediately after ablation and persisting for $\approx$ 2 $\mu$s, which is also consistent with detection of an emission signal.  No dependence on laser wavelength was observed for this prompt signal detected in the absorption. 
 
\begin{figure}[hbt!]
\includegraphics[width=0.8\linewidth]{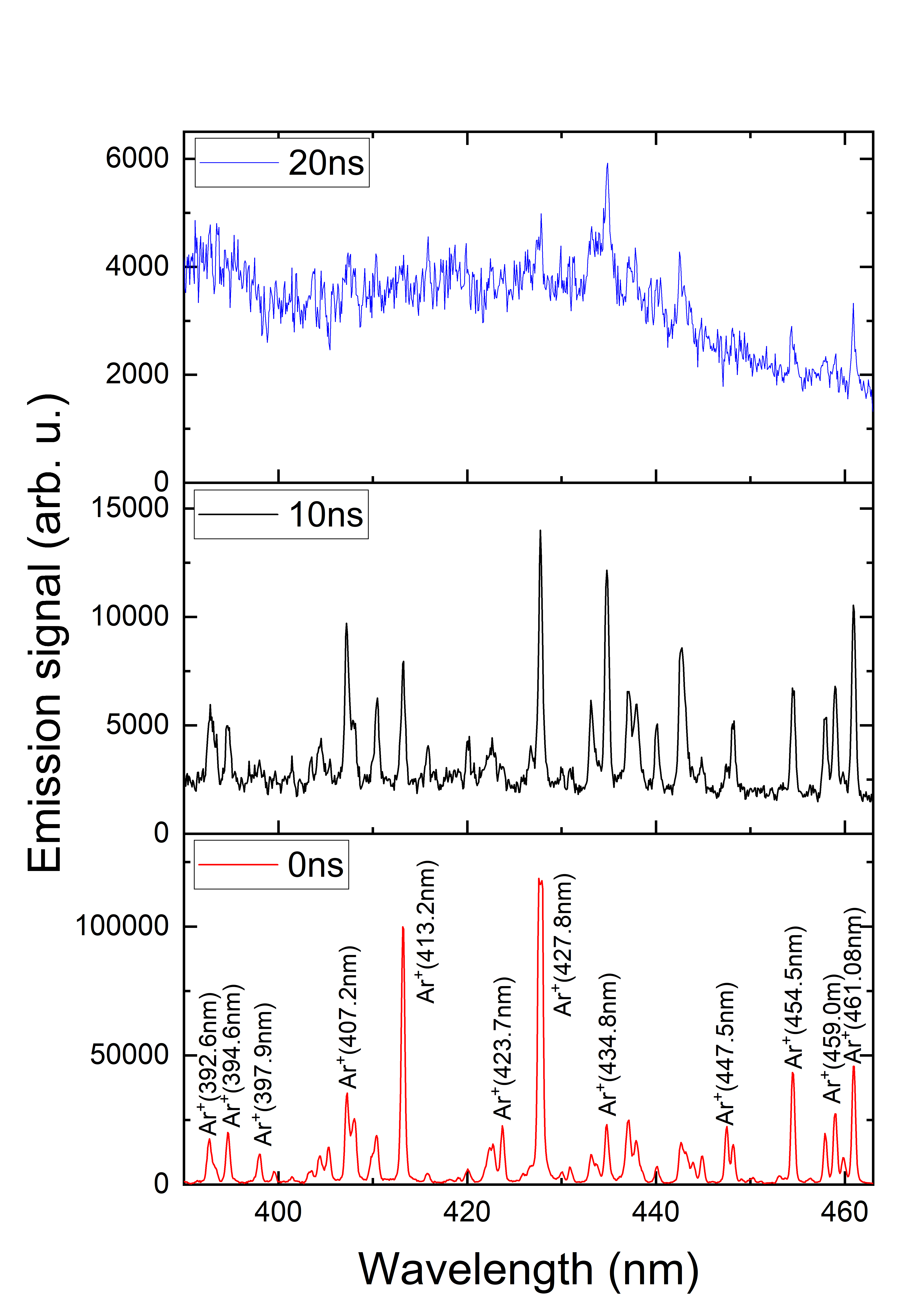}
\caption{\label{ambient-plasma} The spectral features recorded at 6 mm at various times show ambient gas excitation and ionization. A gate width of 3 ns was used for the measurement. The time delays given (0 ns, 10 ns and 20 ns) correspond to the time after the peak of the laser pulse.}
\end{figure}

 The prompt signature observed in both emission and absorption immediately after ablation in (Fig. \ref{EA-TOF-0-2us}) is observed to be nearly independent of distance from the target.  In contrast to ablation under vacuum conditions, the presence of an ambient gas prevents free expansion of atoms from the target, and thus it is highly unlikely the prompt signal arises from atoms ejected from the sample surface.  Instead, previous reports documented the existence of an ambient gas plasma when an LPP is generated in  an ambient gas medium.\cite{2006-Hari-JAP, PoP2019-Ding, Riju-APL} To investigate this further, emission features were collected in the spectral window 390-465 nm at a distance of 6 mm from the target at very early times and results are given in Fig.\ref{ambient-plasma}. The gate width of the measurement was 3 ns and the delay times used were 0 ns, 10 ns and 20 ns after the peak of the laser pulse. The spectral features showed intense emission from Ar$^+$ lines immediately after the laser-plasma onset and the prominent Ar$^+$ lines are marked in the spectrum. The Ar$^+$ spectral intensity decays rapidly and at 20 ns the collected spectrum shows a broad continuum-like emission.   The spectrum contains no Al lines from the plasma generated at the target, which will reach the observation window at later times. These results indicate that when an LPP is generated in an ambient gas medium, the plasma from the target may be preceded by excitation of an ambient gas plasma.  There exists a controversy on explaining the mechanisms leading to this prompt excitation and ionization of ambient gas. Some of the suggested mechanisms are excitation by the prompt electrons generated by the plasma\cite{Riju-APL,Amoruso1999-prompt} and photo-ionization caused by the high energy photons (e.g., VUV emission)\cite{Ratynskaia-APA-2014-prompt}.

\subsection{Temporal evolution 0-10 $\mu$s}
\begin{figure}[hbt!]
\includegraphics[width=0.8\linewidth]{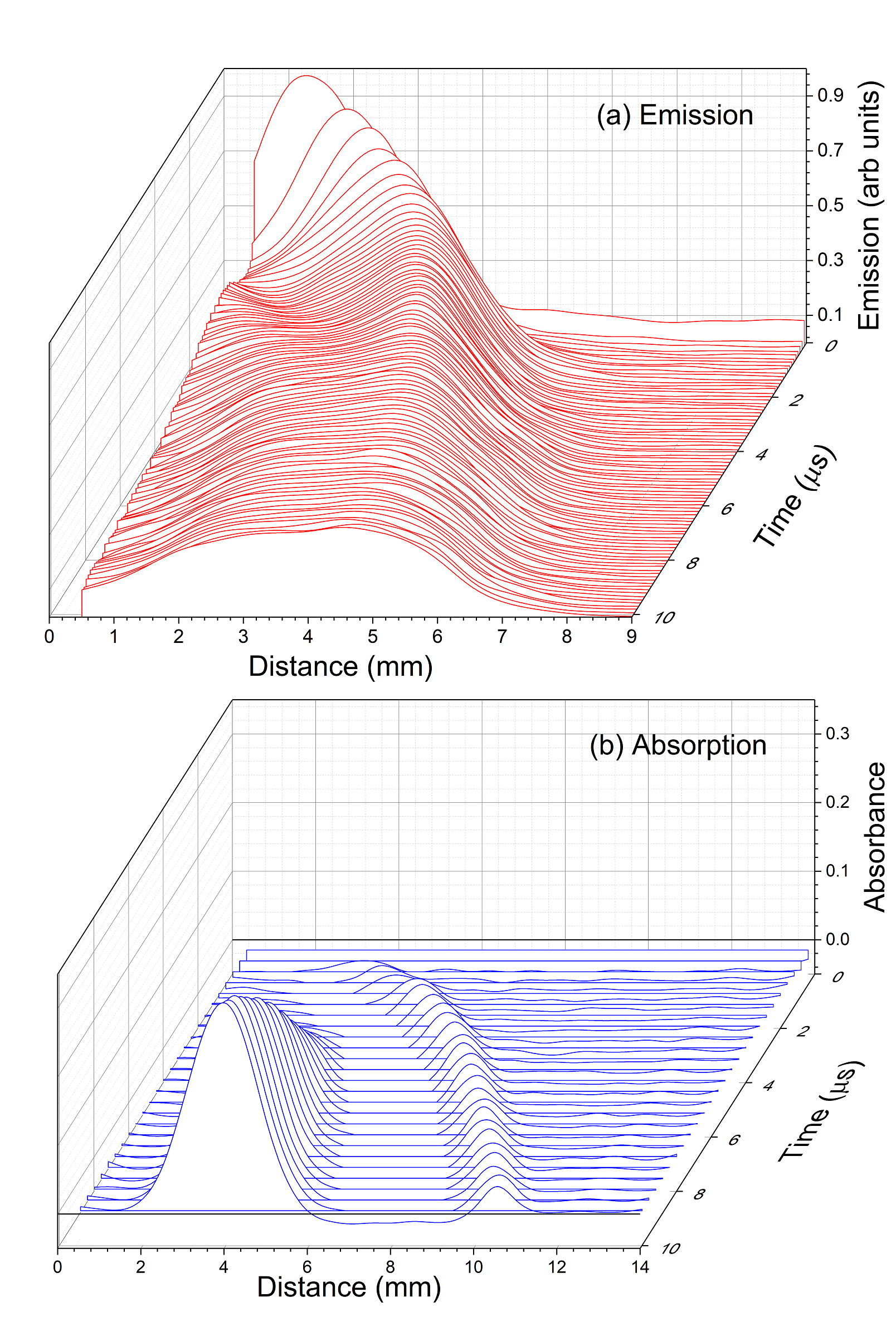}
\caption{\label{EA-TOF-0-10us} The spatio-temporal evolution of emission and absorption signals for times $\leq 10 \mu$s after the laser-plasma onset.}
\end{figure}

 The spatio-temporal evolution of emission and absorption profiles recorded at times $\leq 10 \mu$s after the laser-plasma onset are given in Fig. \ref{EA-TOF-0-10us}. Emission profiles showed that the spatial extent of plume from the target is $\approx$ 8 mm within 10 $\mu$s. The peak emission signal decays and the spatial profile expands during this time period as well.  An apparent reduction in emission signal is visible near the center of the plume profile which become less pronounced over time.
  
 The absorption TOF shows the shock-wave feature continuing to propagate away from the target, with a decreasing velocity.  A negative absorption region is visible behind the shock-front.  From 2-10  $\mu$s, a spatially localized absorption signal develops which is centered near 4 mm and clearly distinct from the propagating shock-wave feature.  During this time period, the Al absorption spectrum was too weak for reliable fitting and so fit parameters are not shown.  However, effects of the shock-wave were visible in the absorption spectral data and were consistent with observations from the TOF data.

\begin{figure}[hbt!]
\includegraphics[width=\linewidth]{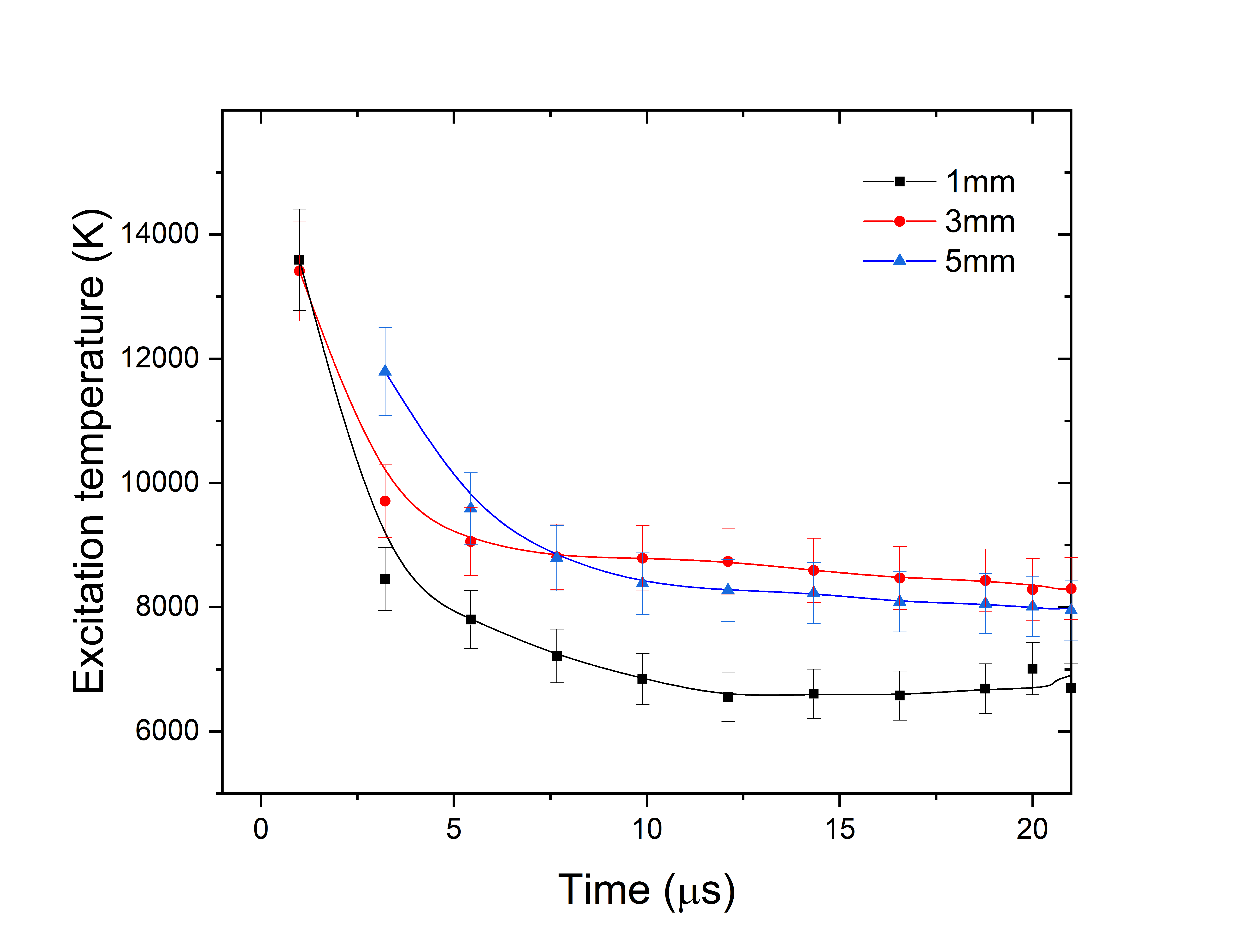}
\caption{ \label{ETemperature} The time evolution of excitation temperature measured at various spatial points in the plasma is given.}
\end{figure}

The lack of absorption for $t \leq$ 2 $\mu$s while strong emission is observed is consistent with conditions of high excitation temperature.  Fig. \ref{ETemperature} shows the excitation temperature determined from OES.  The measured excitation temperature is $\approx$ 14,000 K at 1 $\mu$s after ablation and decays to $\approx$ 7000-9000 K within 10 $\mu$s.  Due to the considerations presented in Fig. \ref{Boltzmann}, the decrease in excitation temperature over this time period is also expected to result in decreasing emission and increasing absorption signals, consistent with the TOF results.

The similar spatial profiles observed in both emission and absorption for t $ \sim$ 2-10 $\mu$s indicate that over this time period the plume is spatially offset from the sample surface, i.e. with a lower Al number density near the target.  The similar profiles observed in absorption and emission also rules out high excitation temperature as a primary cause for the lower absorption near the surface.  The spatial dependence of the absorption and emission may indicate an increased atomic density in the boundary region between the expanding plume and the ambient gas (due to the snowplow effect).

\subsection{Temporal evolution 10 $\mu$s - 20 $\mu$s}
 
\begin{figure}[hbt!]
\includegraphics[width=0.8\linewidth]{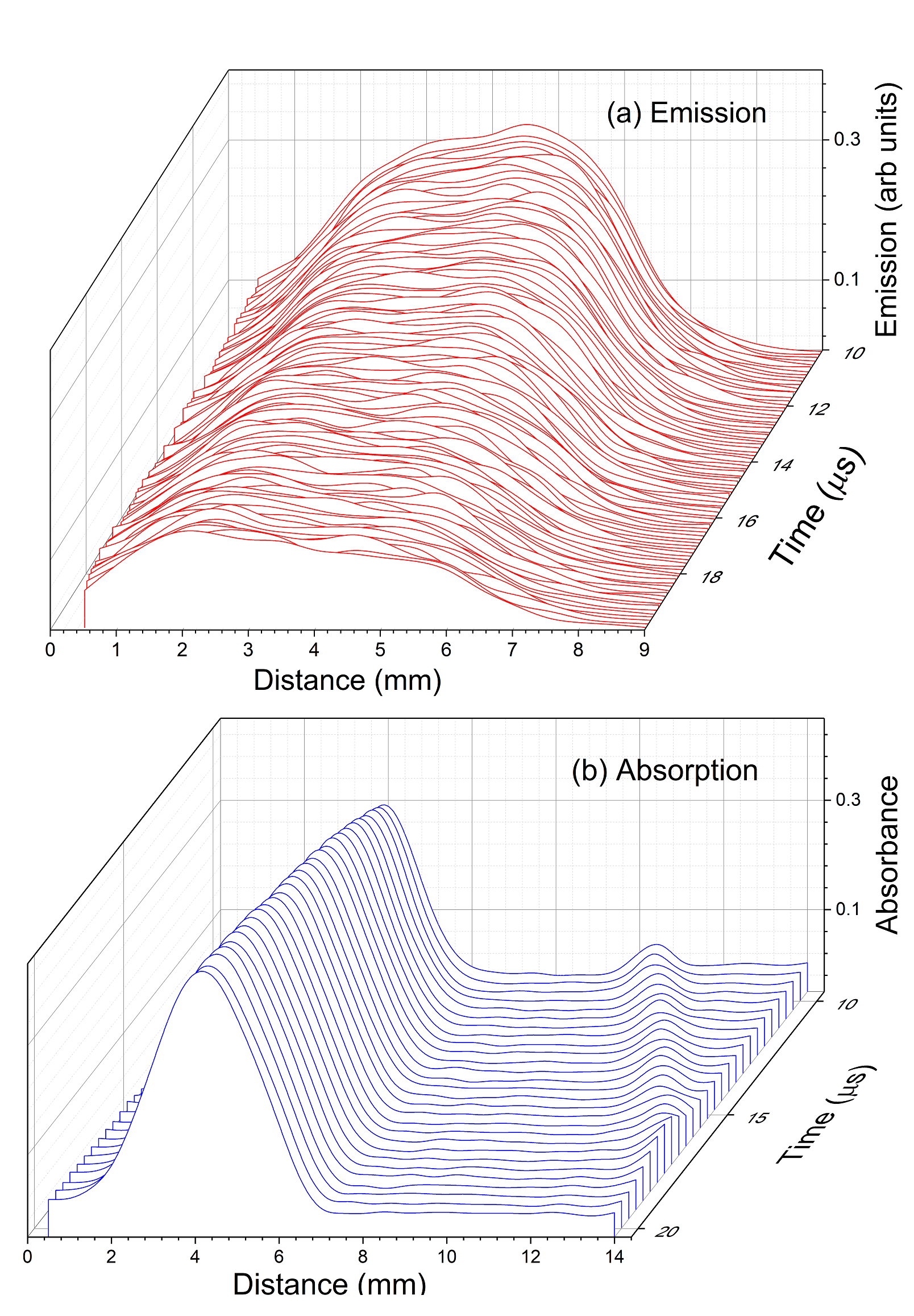}
\caption{\label{EA-TOF-10-20us} The spatio-temporal evolution of (a) emission and (b) absorption signals for times between 10 $ \mu$s and 20 $ \mu$s.}
\end{figure}

\begin{figure}[hbt!]
\includegraphics[width=0.8\linewidth]{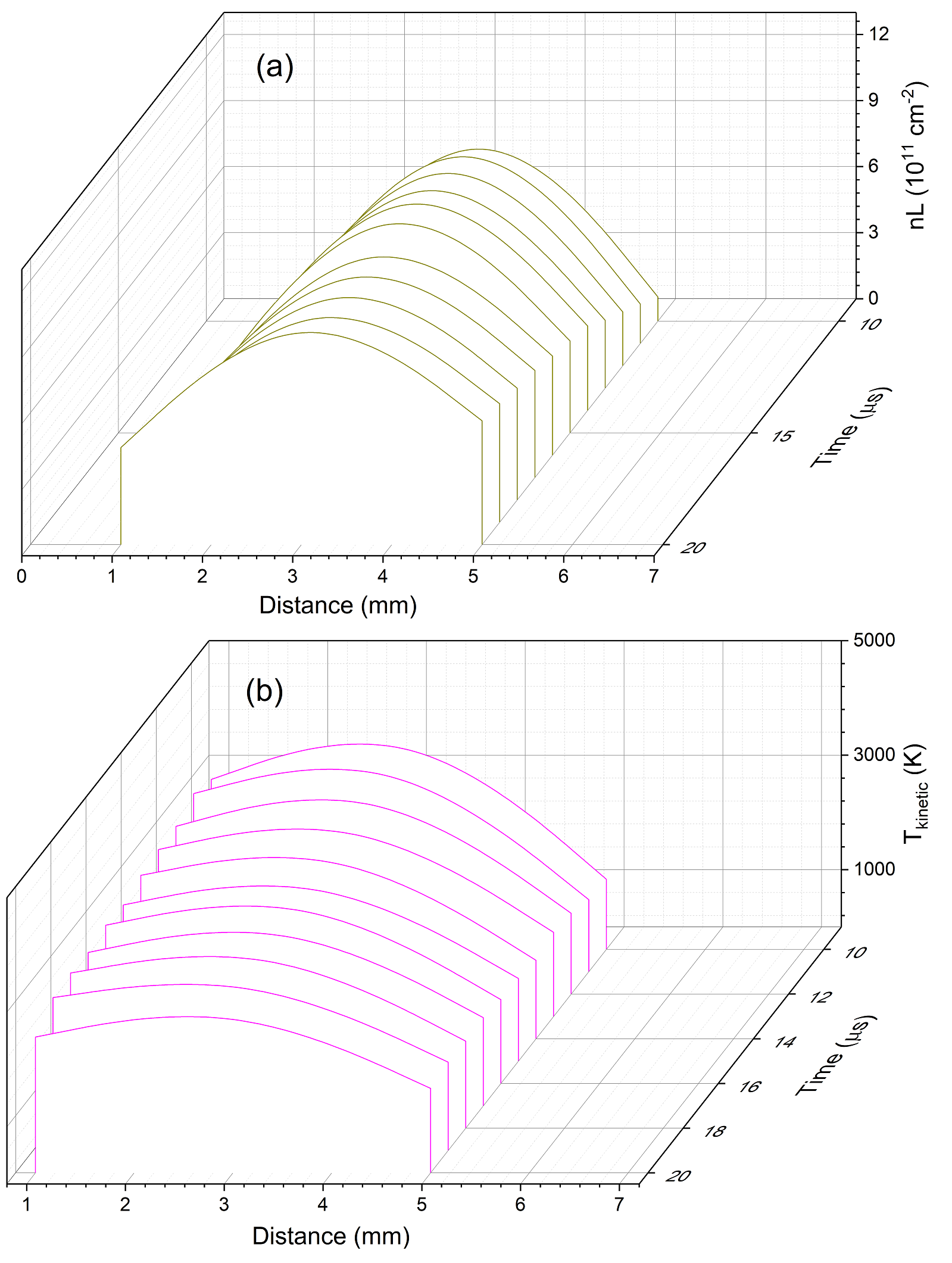}
\caption{\label{NT-10-20us} The spatio-temporal evolution of (a) column density and (b) kinetic temperature measured from the LAS spectral fit parameters are given for times between 10 $ \mu$s and 20 $ \mu$s  after the laser-plasma onset.}
\end{figure}

 The spatio-temporal evolution of emission and absorption TOF profiles recorded at times between 10 $\mu$s and 20 $\mu$s after the laser-plasma onset is given in Fig.\ref{EA-TOF-10-20us}. The emission and absorption profiles from the plume remain centered at a distance offset from the target surface. The absorption TOF profiles show that the shock wave feature leaves the measurement region in this time period. Near the target surface, the absorption spectra become strong enough for reliable fits during this time period, and Fig.\ref{NT-10-20us} shows the column density and kinetic temperatures determined from the fit parameters.       
 
The column density results confirm that the plume is spatially located away from the sample surface, and  the highest kinetic temperatures are measured  near the plume center.  Qualitatively, the spatial profiles of the kinetic temperature from LAS in Fig. \ref{NT-10-20us} match those of the excitation temperature from OES shown in Fig. \ref{ETemperature}, which also shows a higher temperature at 3 mm versus 1 mm and 5 mm.   However, the magnitude of measured excitation temperatures are significantly higher than the measured kinetic temperatures.  As discussed in prior references \cite{Aguilera2004, 2021-SCAB-Nicole}, this behavior is expected due to the spatial integration along the measurement path and differences in measurement of emission versus absorption.  Specifically, the emission measurement is weighted toward spatial regions of higher temperature (where the emission signal is stronger), whereas the absorption measurement is weighted toward spatial regions of colder temperature (where the absorption signal is stronger).  Thus, a measurement of excitation temperature from emission is expected to be higher than a corresponding measurement of kinetic temperature from absorption, especially during early times of plasma evolution when large spatial gradients are present.  Both emission and absorption measurements are intrinsically path-integrated, and hence the differences in temperature along the line-of-sight provide an effective “weighting” to the measurement via the level number densities. 
 
\begin{figure}[hbt!]
\includegraphics[width=0.9\linewidth]{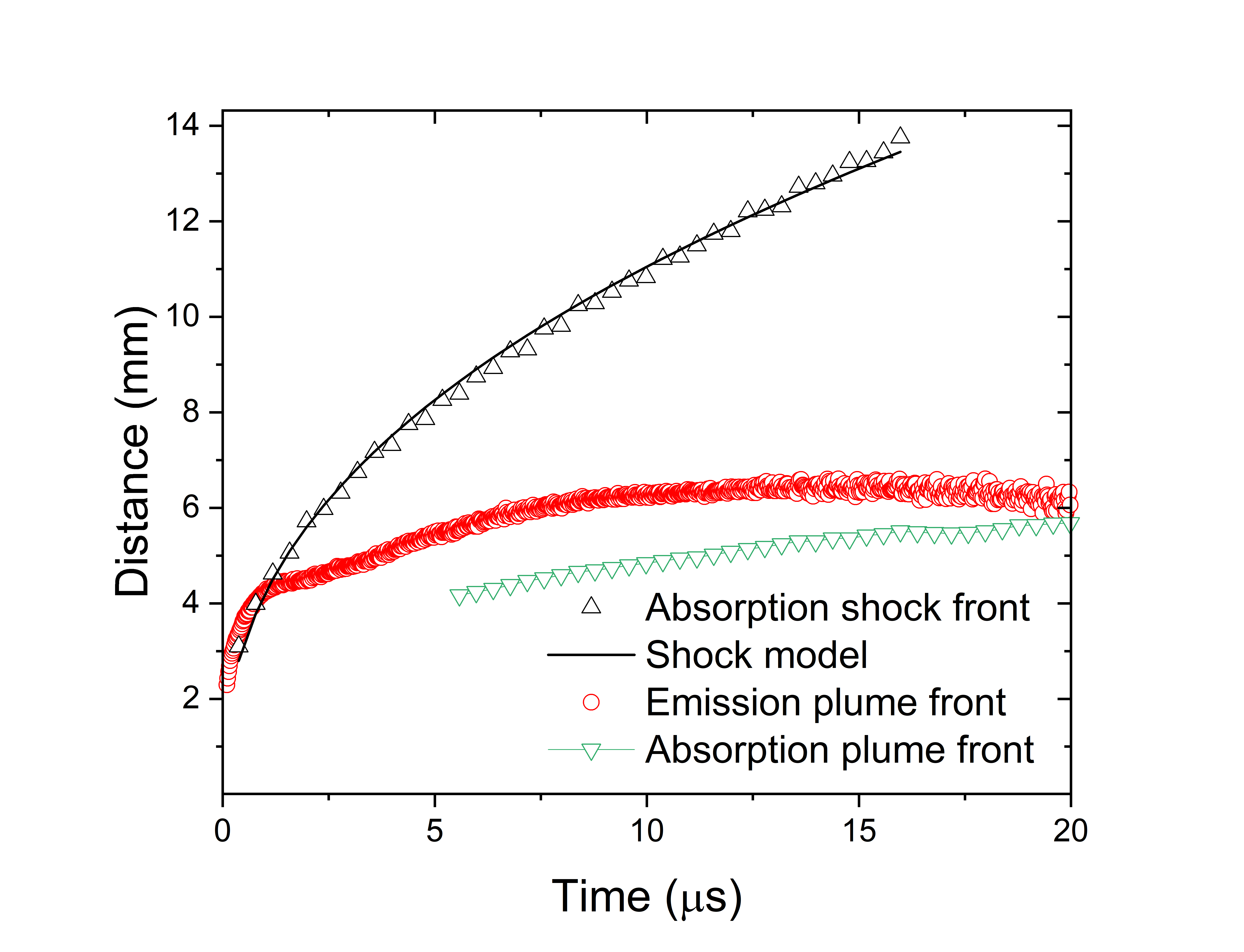}
\caption{\label{RT-plot} The distance-time (R-t) plot for the absorption shock, the emission plume front, the absorption plume front are given. The smooth black line represents the  R $\propto $ t$^{0.42} $ fit.}  
\end{figure}

 Fig. \ref{RT-plot} shows results from additional analysis of the expanding plume, plotted as a distance-time (R-t) plot for the shock wave and plume front components obtained from emission and absorption TOF. Here, the plume front is defined as the 50\% of the maximum signal intensity at each time step.  The location of the peak for the shock wave feature in the absorption TOF data is plotted versus time, along with a fit to a shock model.  Due to the experiment being performed in 34 Torr Ar ambient gas, a strong shock wave is expected to form during the LPP expansion, and the good agreement between the experiment and model confirms this to be the case.  The observed transient increase in absorption as the shock wave propagates through the laser beam path could be due to beam steering effects, reducing the amount of light reaching the detector.  Likewise, the negative absorption peak behind the shock wave may also be due to beam steering or lensing effects, causing the light reaching the detector to increase. Typically, the shock waves moving outward into the ambient from an LPP is accompanied by a rarefaction wave moving inward (towards the target).  However, we note that the exact reasons for the observation of negative absorption signal is unknown at this point and will require additional study.
 
Fig. \ref{RT-plot} also shows the positions of the expanding plume front obtained from the emission and absorption signals.   Before $\approx$ 2 $\mu$s, the shock waves recorded using absorption TOF propagates together with the emission plume front and after $\approx$ 2 $\mu$s the shock front is detached and propagates away from the emission plume front. Similar observations were made on laser-produced air detonation plasma where the plasma and shock waves expanded with similar velocities until $\approx $1 $\mu$s and then the shock wave became decoupled to the plasma and expanded away.\cite{2015-POP-Hari-air}  The absorption plume front appears to lag behind the emission plume front, although this may reflect differences in absorption versus emission strength due to excitation temperature effects.  Nevertheless, in both absorption and emission the plume front appears to stop expanding at a distance of $\approx$ 6 mm from the sample, and at a time between 10-20 $\mu$s.  LPPs typically expand very rapidly at early times immediately after the laser-plasma onset regardless of the ambient pressure due to large plasma ram pressure, which could be $\geq$1000 atm,\cite{2012-POP-Hari} and the plume front decelerates later when it interacts with the ambient gas medium.  Hence the deceleration seen the emission and absorption fronts at later times are due to plasma confinement.

\subsection{Temporal evolution 0-100 $\mu$s}

\begin{figure}[hbt!]
\includegraphics[width=0.8\linewidth]{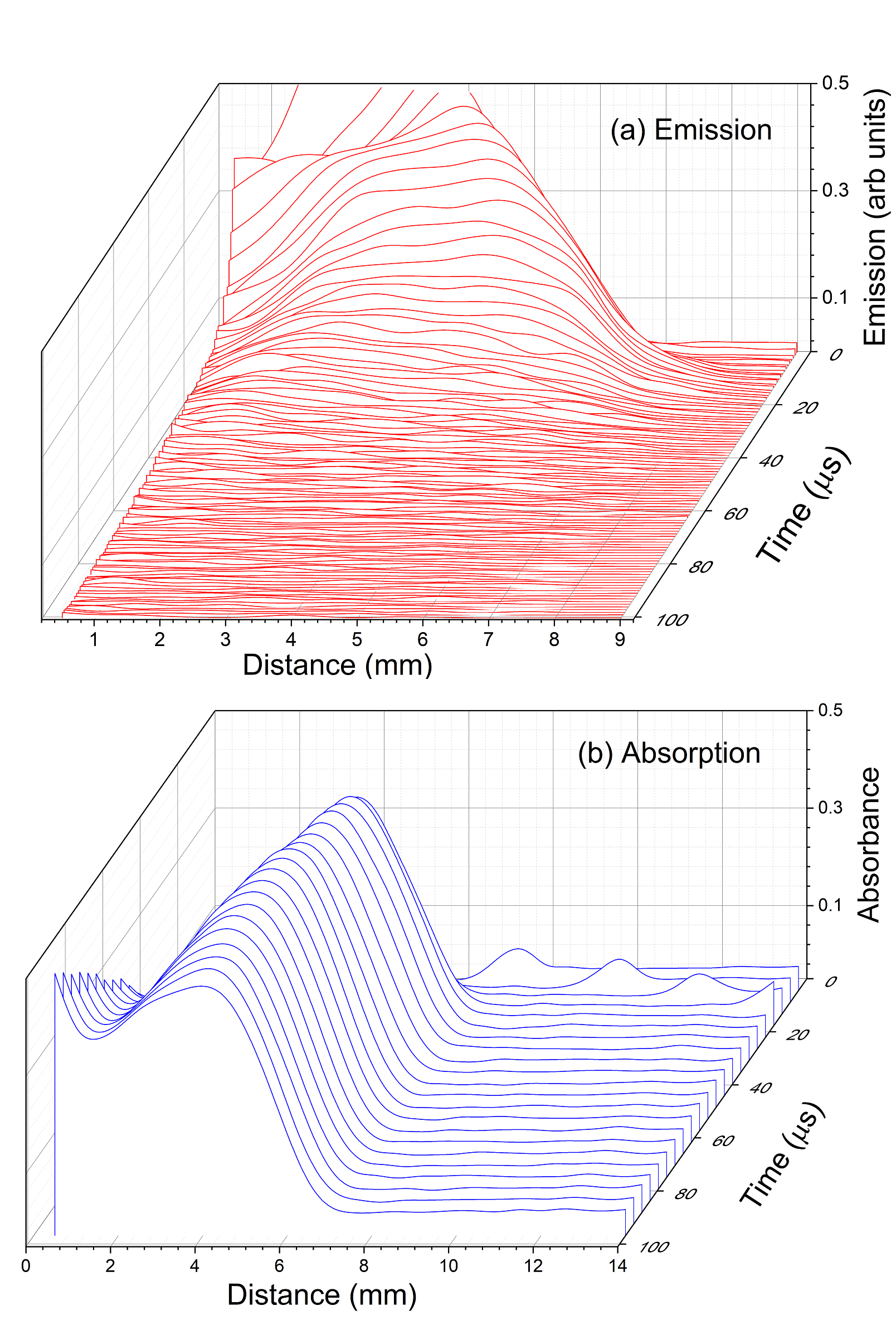}
\caption{\label{EA-TOF-0-100us} The spatio-temporal evolution of emission and absorption signals for times 0- 100 $ \mu$s  after the laser-plasma onset are given.}
\end{figure}

Spatio-temporal evolution of emission and absorption profiles recorded at times up to 100 $\mu$s after the laser-plasma onset are given in Fig.\ref{EA-TOF-0-100us}.
The emission signal extends from the target to $\approx$ 8 mm and persists to $\approx$ 100 $\mu$s. The emission signal strength decreases from 0-40 $\mu$s.  By comparing with the absorption signal, the reduction in emission signal is not driven by a  decrease in number density, as shown by nearly constant absorption over this time period. Instead, the decrease in emission strength is driven by the reduction in excitation temperature over this time period. The absorption signal shows  a spatially localized signal centered near $\approx$ 4 mm and extending to $\approx$ 7 mm from the target. A second absorption component near the target appears at $\approx$ 20 $\mu$s and grows in amplitude beyond 100 $\mu$s; this feature is not apparent in the emission signal.  

\begin{figure}[hbt!]
\includegraphics[width=0.8\linewidth]{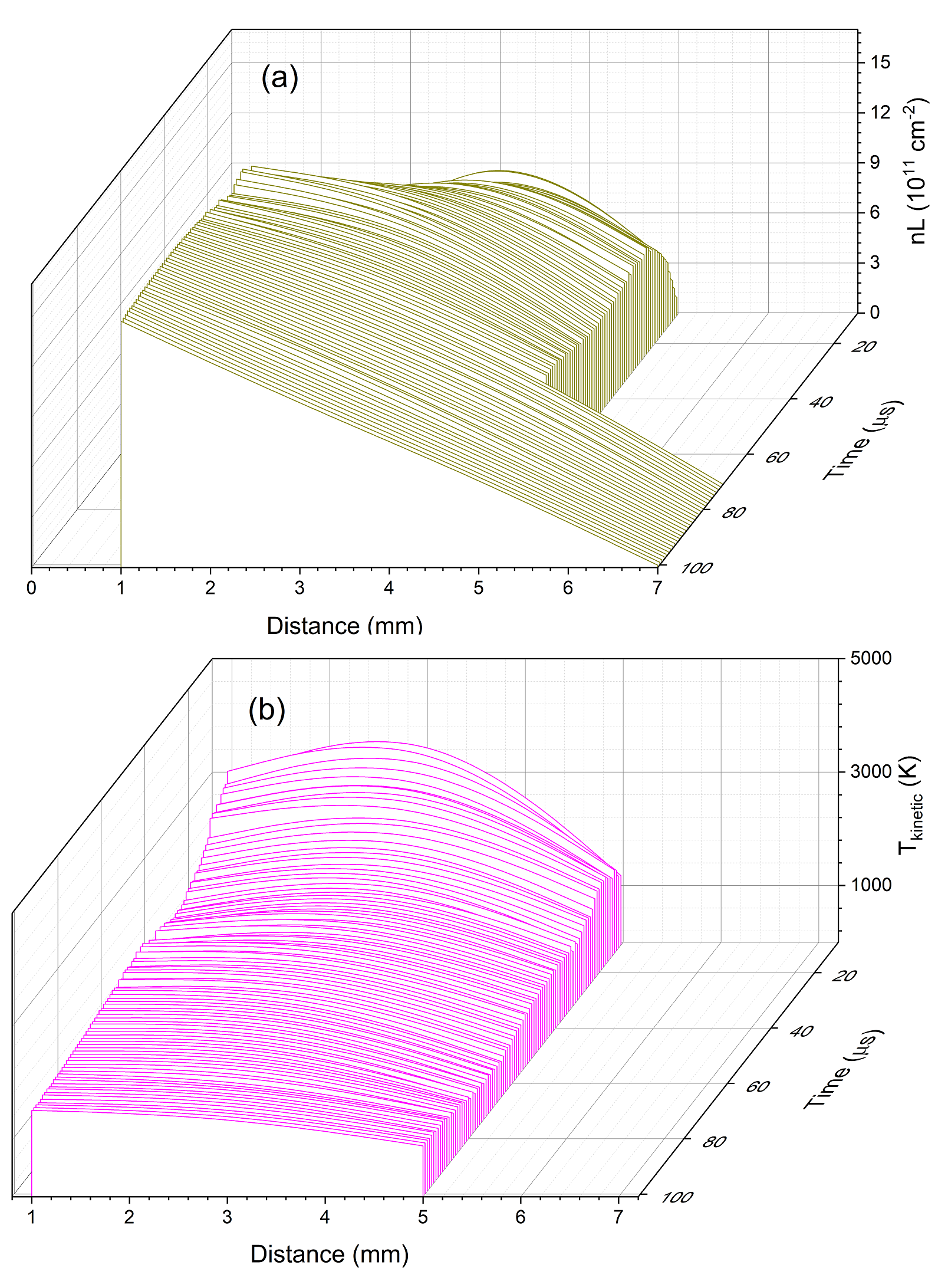}
\caption{\label{NT-10-100us} The spatio-temporal evolution of (a) column density and (b) kinetic temperature are given for times $\leq$ 100 $ \mu$s.}
\end{figure}

The spatio-temporal evolution of column density and kinetic temperature for times $\leq$ 100 $\mu$s are given in Fig. \ref{NT-10-100us}. The column density determined from LAS shows a notable difference from the absorption TOF signals.  In particular, for t $\geq$ 40 $\mu$s the column density is highest near the sample surface and shows a monotonic decay with distance away from the sample.  In contrast, the TOF absorption signal continues to show a peak centered near $\approx$ 4 mm, with an apparent second peak increasing near the sample surface.  This apparent discrepancy can be resolved by considering that the TOF absorption measures the absorption signal at the peak of the Al I spectral profile. As shown in Fig. \ref{FWHM}, the spectral width decreases significantly over the time period up to 100 $\mu$s.  Therefore, over these time periods the peak height is not indicative of the actual column density (which scales with peak area).  In this situation, the spectral measurement is essential for interpreting the TOF results.  Nevertheless, the TOF results are valuable for showing the overall plume extent especially for spatial regions and times with low absorption signals (where spectral fits cannot be performed reliably).

Due to the large decrease in excitation temperature over this time period, the emission signal is reduced in magnitude and eventually becomes undetectable.  In contrast, the absorption signal from the ground state remains strong over this time period. The column density and kinetic temperature results show that the plume continues to expand over this time period, which leads to the number density and temperatures being highest near the sample surface starting at t $\approx$ 40 $\mu$s.  From 40-100 $\mu$s, the plume continues to expand away from the sample surface and the kinetic temperature continues to decrease.  We note that in Fig. \ref{NT-10-100us}, some points are absent in the kinetic temperature but are present in the column density.  This occurs because the peak area could be obtained from the spectral fits whereas the Gaussian component of the spectral width could not be determined with confidence, due to low SNR at these spatial/temporal points.

\subsection{Temporal evolution 100-1000 $\mu$s}
\begin{figure}[hbt!]
\includegraphics[width=0.8\linewidth]{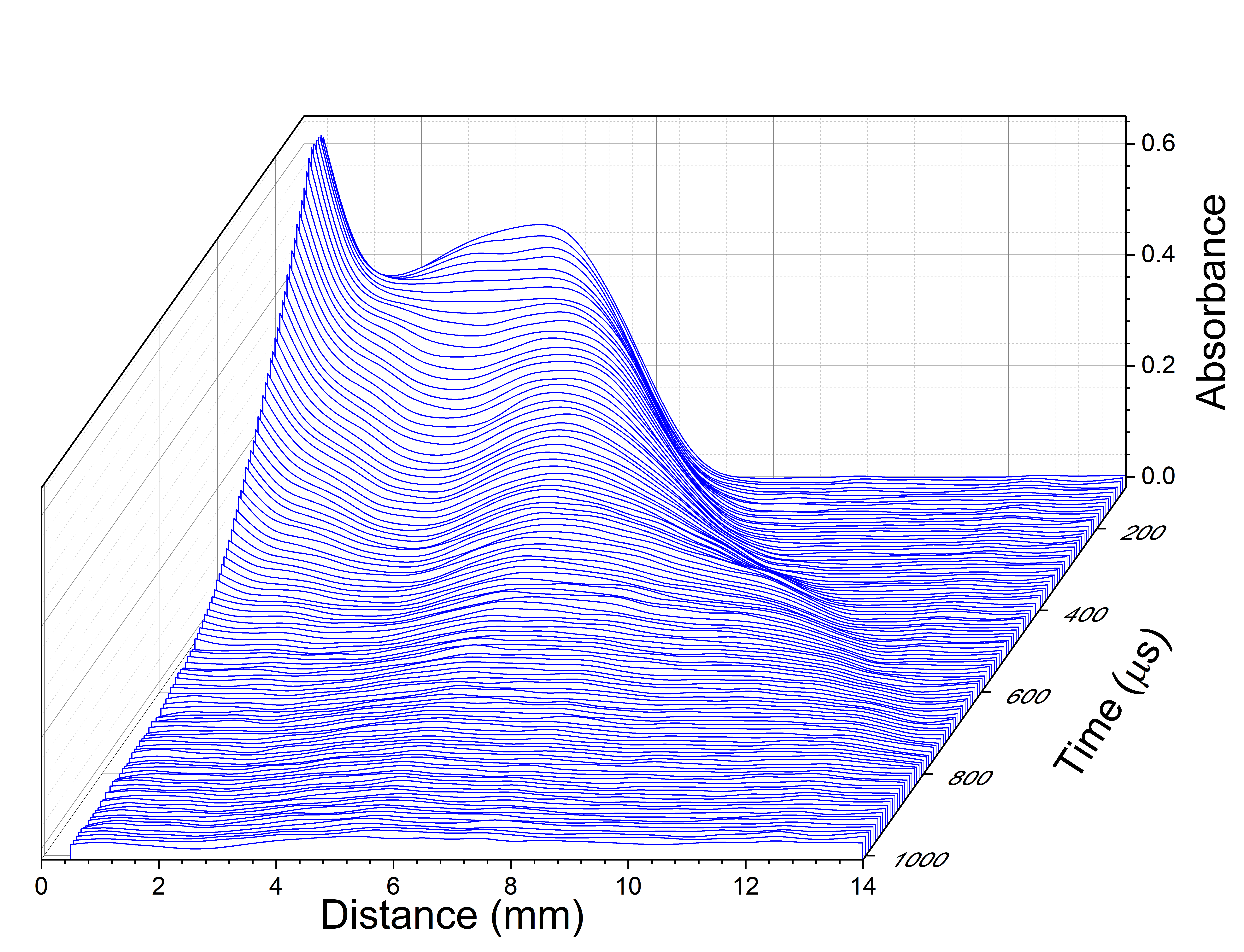}
\caption{\label{EA-TOF-0-1000us} The spatio-temporal evolution of absorption for times 100- 1000 $\mu$s after the plasma onset is given.}
\end{figure}

\begin{figure}[hbt!]
\includegraphics[width=0.8\linewidth]{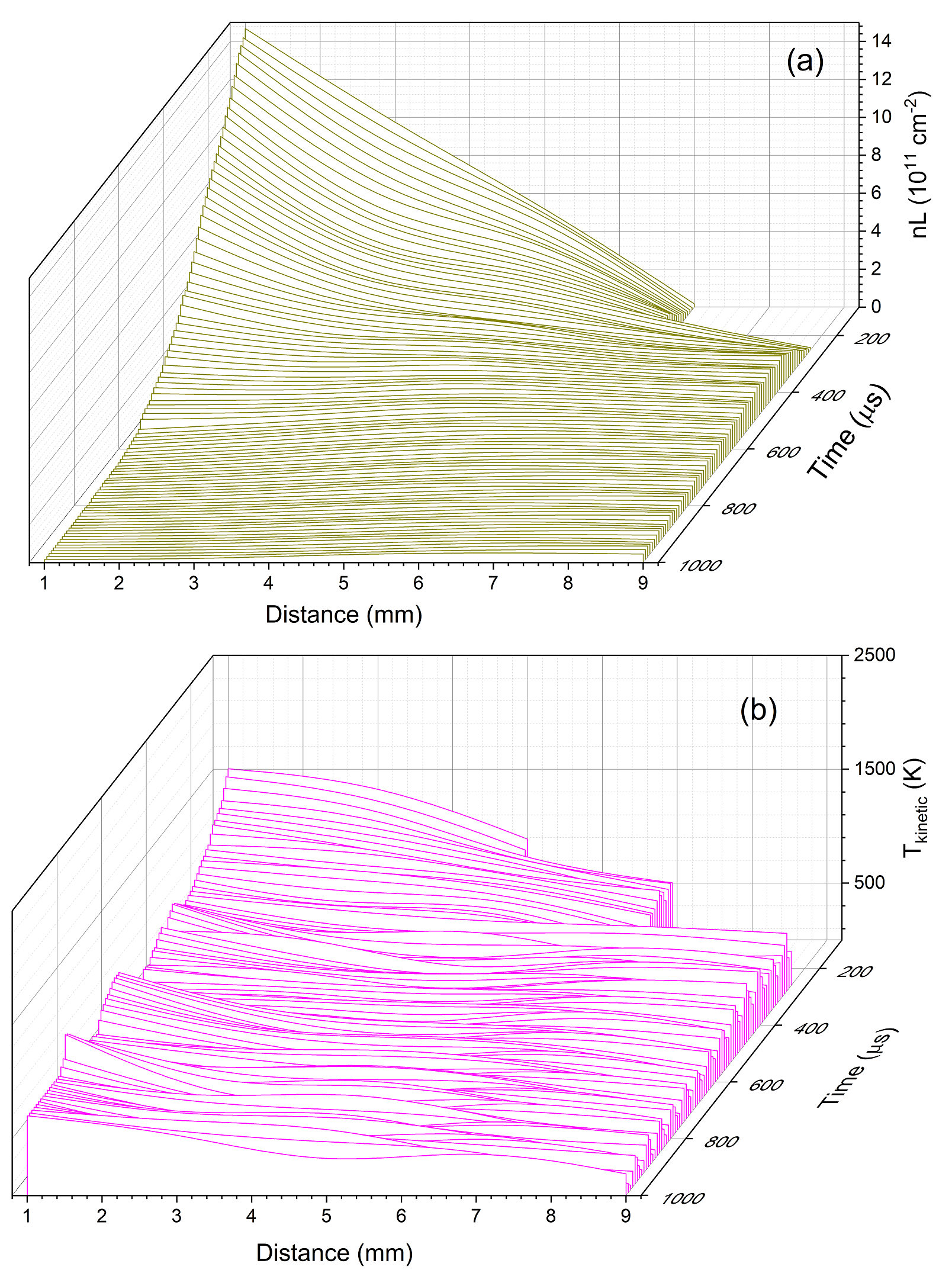}
\caption{\label{NT-100-1000us} The spatio-temporal evolution of (a) column density and (b) kinetic temperature are given for times 100-1000 $ \mu$s.}
\end{figure}

The spatio-temporal evolution of emission for times 100-200 $\mu$s and absorption 100-1000 $\mu$s  after the laser-plasma onset are given in Fig. \ref{EA-TOF-0-1000us}. The spatio-temporal evolution of column density and kientic temperature during this time window are given in Fig. \ref{NT-100-1000us}. The emission signal could not be detected in this time period, due to the low excitation temperatures. The absorption signal continued to evolve with time and the peak center moved outward from the sample. At $\approx$ 500 $\mu$s, the absorption signal near the surface has again disappeared, with a similar behavior observed in the column density.  The spatial profile from absorption and column density then continues to move to larger distances over time, indicating a slow diffusion of the plume away from the target.  The kinetic temperature also continues to decrease over this time period and becomes more uniform with distance.

\begin{figure}[hbt!]
\includegraphics[width=0.8\linewidth]{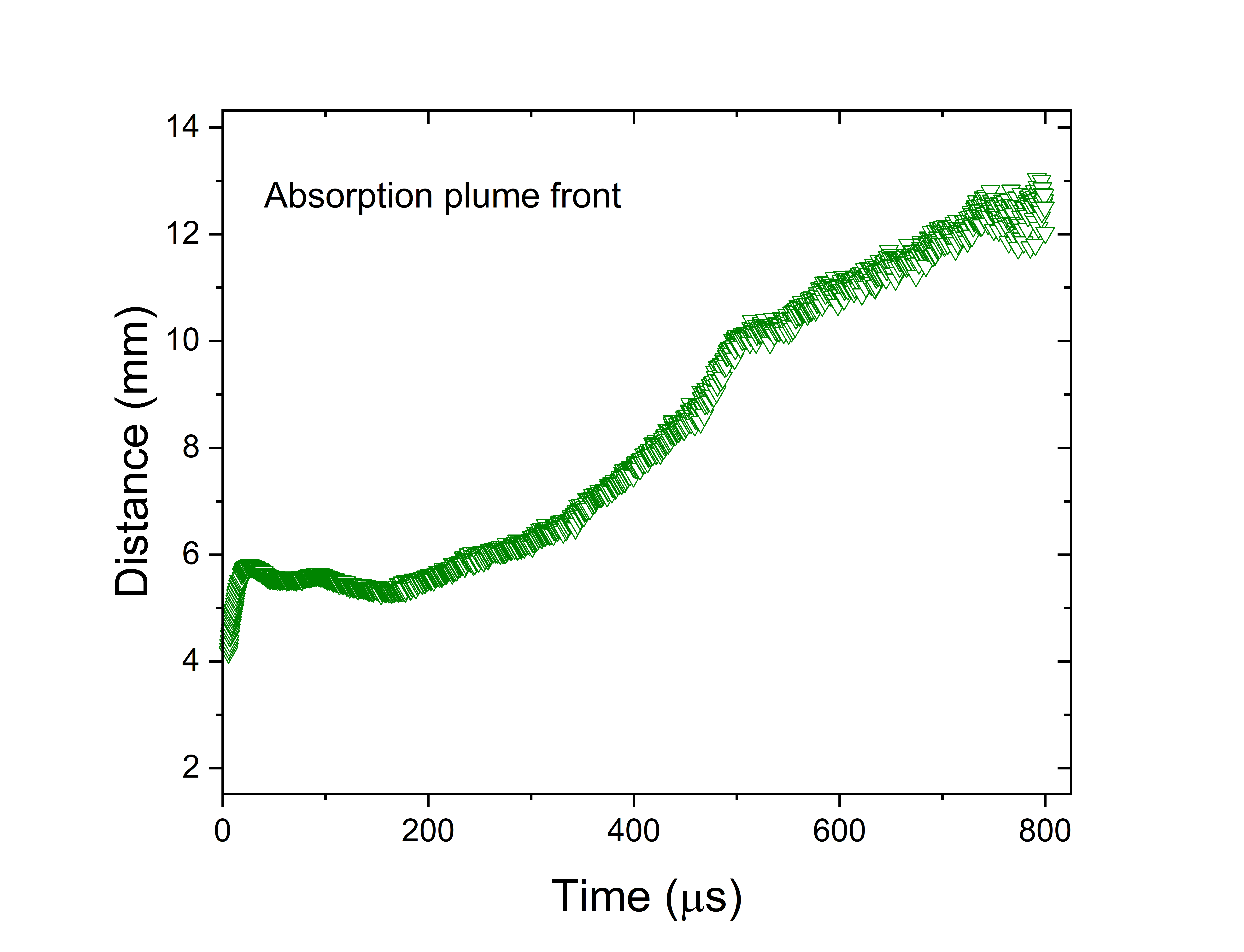}
\caption{\label{Absorption-plume-font} The R-t plot of the absorption signal front at all times during the plume lifetime .}
\end{figure}

The R-t plot of the absorption signal front at all times during the plume lifetime is given in Fig. \ref{Absorption-plume-font} which showed a complex propagation pattern. The absorption plume front exhibits a rapid expansion at early times until $\approx$ 20 $\mu$s, followed by an apparent stagnant region for times between $\approx $ 20-160 $\mu$s). At later times, the absorption plume front continues to propagate away from the target but with a slower velocity.  When an LPP interacts with a cover gas, the plume species will be decelerated until the plume pressure becomes similar to cover gas pressure. Hence, the propagation of the absorption signal front at very late times of plume evolution could be related to shock wave collapse and diffusion of the plasma species into the ambient medium.\cite{2016-AC-Hari} 

\section{\label{sec:conclusion}Summary and Conclusions}

The excited and ground state population kinetics of the Al I (394.4 nm) transition showed several interesting features. Both emission and absorption TOF evolution showed a prompt signature immediately after the laser-plasma onset and this is caused by the ambient gas excitation and ionization. The absorption TOF profiles showed an early time signal which matches well with shock wave propagation. Multiple peaks in their temporal profiles are observed  both in the emission and absorption TOF profiles. The peak absorbance with distance showed two maxima: one closer to the target and the second one at $\approx$ 4 mm away from the target. The peak absorbance signal is also found to be delayed closer distances from the target surface. The difference between the kinetics of absorbance and emission signals were explained based on the temperature dependence of Boltzmann distribution and effects of time-varying spectral absorption linewidths.

The measured FWHM (Fig. \ref{FWHM}) showed an overall decrease with time and distance from the target. The spectral lines in an LPP get broadened due to various mechanisms. Typically the Stark effect, contributed by charged particles, is prominent only in early times of plasma evolution ($\leq$ 2 $\mu$s) and closer distances when the electron densities are higher.\cite{2018-APR-Hari} The Doppler effect,  contributed by the thermal motion of species with respect to the observer, becomes prominent at later times of plasma evolution. Van der Waals pressure broadening contributes to the Lorentzian linewidth, but is relatively small ($\approx$ 10-20 \% of the Gaussian widths) at the low pressures used in these experiments.\cite{2021-PRE-Hari} In all spatial positions studied, the FWHM of the Al transition approaches $\approx $ 3 GHz. The changes in the linewidth is also found to be marginal at later times of plasma evolution $\geq$ 300 $\mu$s and at larger distances from the target ($\geq $ 7 mm).

The spatio-temporal evolution of ground state column density showed a rapid rise and decay at early times. It is important to note that the measured column densities represent the ground level population rather than defining the entire Al atomic number density in the plasma system.  Since the selected transition is a resonance line, a significant population exists at lower temperature plasma conditions. At early times, the plasma excitation temperature is high and the ground state of the Al I transition is depopulated. However, at late times of plume evolution, the excited state population becomes negligible and hence the absorption measurement more closely represents the total atomic column density. We also noticed a marked difference between the spatio-temporal evolution of absorbance signal and the column density at intermediate times. It indicates the TOF absorption profiles recorded by keeping the probe beam at the peak center is not necessarily a true representation of atom density.  The number density  is measured using the peak area which in turn depends on both peak height and width. The varying temperature with time and space may influence the width of the spectral profile. 

Both excitation and kinetic temperatures decayed with time when the measurements were made at shorter distances from the target although rapid decay was observed for excitation temperature.  The reduction in kinetic temperature with time is also noticeable at later times.  When an LPP is generated in vacuum, the temperature of the system will decay rapidly with time.\cite{Schaeffer-JAP2016}  In the present experiment, an ambient gas with a pressure of 34 Torr was present. The presence of the ambient or cover gas reduces the plume expansion and confines the plasma, which effectively reduces the rapid temperature decay seen in vacuum environment.\cite{aragon2008characterization} The kinetic temperature with distance showed a maximum at 3 mm which is consistent with excitation temperature measurements.   

 The measured temperatures using emission and absorption spectroscopy under the assumption of LTE showed differences in magnitudes. In general, LTE is considered a valid assumption in LPPs on the time scales of ~100 ns to several microseconds\cite{cristoforetti2010local}; its validity at late times in plasma evolution has not been conclusively established.  Under the assumption of LTE, the plasma should be collisionally-dominated, and hence, all temperatures should be similar at a  spatial position in the plasma and a certain time during its evolution.  However, such a difference in temperature is not unexpected for an LPP, considering its inhomogeneous nature. It has to be mentioned that the kinetic temperature is measured using spectral profiles of Al I 394 nm, and the excitation temperature was measured using emission features of several Cr I transitions which originated from significantly higher excited levels. Hence, the excitation temperature was determined using emission spectroscopy is expected to be higher from the hotter regions of the plasma. Similarly, the kinetic temperature is measured using neutral Al I, whose absorption will be highest from cooler plasma regions. 
 
 In summary, the present work provided a comprehensive comparison between the spatio-temporal evolution of excited and lower level atom density distributions in an LPP system by combining emission and tunable laser absorption spectroscopy and TOF measurements. The TOF measurements provided information on spatio-temporal evolution of excited state populations (emission) and ground state populations (absorption) over multiple time scales.  Emission from ambient gas excitation was evident, followed by shock-wave propagation and the rapid initial expansion and stagnation of the LPP from the target surface, and finally a slower diffusion of the LPP away from the target surface.  Additional measurements of emission spectra provided information on excitation temperatures at early times of LPP evolution, while analysis of absorption spectra provided information on column densities and kinetic temperatures at later times of LPP evolution.  The spectral measurements and analysis were shown to be essential for interpretation of the TOF results by providing additional information on temperatures and spectral linewidths.  Nevertheless, the TOF measurements were highly efficient for showing the spatio-temporal LPP dynamics with higher spatial and temporal resolution than is practical with full spectral acquisition and fitting.  Furthermore, the TOF data could be measured in time/space regions where the SNR is too low for reliable spectral analysis.  Overall, the combination of absorption and emission TOF and spectral measurements provides a more complete picture of LPP spatio-temporal dynamics than is possible using any one technique alone.

\begin{acknowledgments}
This work was partially supported by the Department of the Defense, Defense Threat Reduction Agency (DTRA) under award number HDTRA1-20-2-0001. The
content of the information does not necessarily reflect the position or the
policy of the federal government, and no official endorsement should be
inferred.  Pacific Northwest National Laboratory is a multi-program national laboratory operated by Battelle for the U.S. Department of Energy under Contract DE-AC05-76RL01830.
\end{acknowledgments}

\section {AUTHOR DECLARATIONS}
\subsection{Conflicts of Interest} 
MCP is a part-time employee of a small business Opticslah, LLC.

\subsection {Data availability} 
The data that support the findings of this study are available
upon reasonable request to corresponding author.

\section*{References}
\bibliography{LAS-Spatial}

\end{document}